\documentclass[ALICE,manyauthors]{cernphprep}

\def\pt {\mbox{$p_{\rm T}$}\xspace}

\def\pbpb {Pb--Pb\xspace}

\def\jpsi {\mbox{J/$\psi$}\xspace}

\newcommand{\lambdaTheta}{\ensuremath{\lambda_{\theta}}}
\newcommand{\lambdaPhi}{\ensuremath{\lambda_{\phi}}}
\newcommand{\lambdaThetaPhi}{\ensuremath{\lambda_{\theta\phi}}}

\usepackage{datetime}
\usepackage{amsmath}
\usepackage{xcolor}
\usepackage[comma,square,numbers,sort&compress]{natbib}
\usepackage{hyperref}
\usepackage{lineno}
\usepackage{multirow}
\usepackage{gensymb}
\usepackage{graphicx}
\usepackage{placeins}
\usepackage[T1]{fontenc}

\begin{document}%

\begin{titlepage}
\PHyear{2020}
\PHnumber{92}       
\PHdate{19 May}  

\title{First measurement of quarkonium polarization \\ in nuclear collisions at the LHC}
\ShortTitle{Quarkonium polarization in nuclear collisions at the LHC}  

\Collaboration{ALICE Collaboration\thanks{See Appendix~\ref{app:collab} for the list of collaboration members}}
\ShortAuthor{ALICE Collaboration} 


\begin{abstract}
The polarization of inclusive \jpsi and $\Upsilon(1{\rm S})$ produced in \pbpb collisions at $\sqrt{s_{\rm{NN}}}=5.02$ TeV at the LHC is measured with the ALICE detector. The study is carried out by reconstructing the quarkonium  through its decay to muon pairs in the rapidity region $2.5<y<4$ and measuring the polar and azimuthal angular distributions of the muons. The polarization parameters $\lambda_{\theta}$, $\lambda_{\phi}$ and $\lambda_{\theta\phi}$ are measured in the helicity and Collins-Soper reference frames, in the transverse momentum interval $2<p_{\rm T}<10$ GeV/$c$ and $p_{\rm T}<15$ GeV/$c$ for the \jpsi and $\Upsilon(1{\rm S})$,  respectively.
The polarization parameters for the \jpsi are found to be compatible with zero, within  a maximum of about two standard deviations at low $p_{\rm T}$, for both reference frames and over the whole $p_{\rm T}$ range. The values are compared with the corresponding results obtained for pp collisions at $\sqrt{s}=7$ and 8 TeV in a similar kinematic region by the ALICE and LHCb experiments. Although with much larger uncertainties, the polarization parameters for $\Upsilon(1{\rm S})$ production in \pbpb collisions are also consistent with zero.

\end{abstract}
\end{titlepage}
\setcounter{page}{2}
%
%
\section{Introduction}

Quarkonia, bound states of charm (c) and anticharm ($\overline{\rm c}$) or bottom (b) and antibottom ($\overline{\rm b}$) quarks, represent an important tool to test our understanding of quantum chromodynamics (QCD), since their production process involves both perturbative and non-perturbative aspects.
At high energy, the creation of the heavy quark-antiquark pair is a  process that can be described using a perturbative QCD approach, due to the large value of the charm and bottom quark masses ($m_{\rm c}\sim 1.3$ GeV/$c^2$, $m_{\rm b}\sim 4.2$ GeV/$c^2$~\cite{Tanabashi:2018oca}). However, the subsequent formation of the bound state is a non-perturbative process that can be described only by empirical models or effective field theory approaches. Among those, models based on Non-Relativistic QCD (NRQCD)~\cite{Bodwin:1994jh} give the most successful description of the  production cross section, as measured at high-energy hadron colliders (Tevatron, RHIC, LHC)~\cite{Acharya:2017hjh,Acharya:2019lkw,Aaij:2015rla,Sirunyan:2017qdw,Aad:2015duc,Acharya:2019iok,Adam:2019mrg,Abe:1997jz,Abachi:1996jq,Aaij:2018pfp,Adam:2015rta,Aad:2012dlq}.
In the NRQCD approach, the non-perturbative aspects are parameterized via long-distance matrix elements (LDME), corresponding to the possible intermediate color, spin and angular momentum states of the evolving quark-antiquark pair. The values of LDMEs need to be fitted on a subset of the available measurements and can be then considered as universal quantities, in the sense that they can be used in the calculation of production cross sections and other observables corresponding, for example, to different collision systems and energies.
 Other theory approaches, as the Color Singlet Model~\cite{Einhorn:1975ua}, the Color Evaporation Model~\cite{Gluck:1977zm} and the $k_{\rm T}$-factorization~\cite{Baranov:2016clx} are also used to describe the quarkonium production process.

Among the various charmonium states, the \jpsi meson, with quantum numbers $J^{\rm {PC}}=1^{--}$, was the first to be discovered.  It is surely the most studied, also due to the sizeable decay branching ratio to dilepton pairs (($5.961\pm0.033$)\% for the $\mu^+\mu^-$ channel~\cite{Tanabashi:2018oca}) that represents an excellent experimental signature. While the J/$\psi$ production cross sections are well reproduced by NRQCD-based models, it was soon realized that describing the measured polarization of this state represents a much more difficult problem~\cite{Braaten:1999qk}.
The polarization, corresponding to the orientation of the particle spin with respect to a chosen axis,
 can be accessed via a study of the polar ($\theta)$ and azimuthal ($\phi$) production angles, relative to that  axis, of the two-body decay products in the quarkonium rest frame. Their angular distribution $W(\theta,\phi )$ is parameterized as 

\begin{eqnarray}
W(\theta,\phi ) &\propto &\frac{1}{3+\lambda_{\theta}} \left( 1 + \lambda_{\theta}\cos^2{\theta} + \lambda_{\phi}\sin^2{\theta} \cos {2\phi} + \lambda_{\theta\phi} \sin {2\theta}\cos{\phi} \right),
\label{eq:1}
\end{eqnarray}

with the polarization parameters $\lambda_{\theta}$, $\lambda_{\phi}$ and $\lambda_{\theta\phi}$ corresponding to various combinations of the elements of the spin density matrix of J/$\psi$ production~\cite{Faccioli:2010kd}. In particular, the two cases ($\lambdaTheta=1$, $\lambdaPhi=0$, $\lambdaThetaPhi=0$) and  ($\lambdaTheta=-1$, $\lambdaPhi=0$, $\lambdaThetaPhi=0$) correspond to the so-called transverse and longitudinal polarizations, respectively. At leading order, the high-$p_{\rm T}$ production is dominated by gluon fragmentation and therefore the J/$\psi$ would be expected to be transversely polarized~\cite{Braaten:1999qk}. However, the results from the CDF experiment at Tevatron showed that the J/$\psi$ exhibits a very small polarization~\cite{Affolder:2000nn,Abulencia:2007us}, an observation which was impossible to reconcile with the NRQCD prediction. As of today, on the experimental side, accurate results on inclusive and prompt (i.e., removing contributions from b-quark decays) J/$\psi$ polarization have become available at LHC energies~\cite{Abelev:2011md,Acharya:2018uww,Aaij:2013nlm,Chatrchyan:2013cla}. They confirm that this state shows little or no polarization in a wide rapidity (up to $y=4.5$) and transverse momentum region (from 2 to 70 GeV/$c$), with the exception of the LHCb measurements at $\sqrt{s}=7$ TeV~\cite{Aaij:2013nlm}, where the value $\lambda_\theta=-0.145\pm0.027$, corresponding to a weak longitudinal polarization, was obtained in the interval $2<p_{\rm T}<15$ GeV/$c$ and $2<y<4.5$, in the helicity frame (its definition will be given later in Sec. 3).
On the theory side, a huge effort was pursued in order to move to a complete next-to-leading order (NLO) description of the J/$\psi$ production process~\cite{Butenschoen:2010rq,Ma:2010jj}, and to the calculation of the polarization variables~\cite{Butenschoen:2012px,Chao:2012iv}. Further important progress includes a quantitative evaluation of the contribution of feed-down processes (J/$\psi$ coming from the decay of $\chi_{\rm c}$ and $\psi(2{\rm S})$ states) on the polarization observables~\cite{Gong:2012ug}. It was shown that at NLO there are rather large cancellations between contributions corresponding to the different possible combinations of the spin and angular momentum of the intermediate ${\rm c}{\rm\overline c}$ states, reaching a more satisfactory description of the absence of polarization observed in the data~\cite{Feng:2018ukp}. However, those descriptions usually require the inclusion of both cross section and polarization results in the fit of the LDME, leading to a more limited predictive power on the polarization observables and to large variations in the values of the extracted LDME values, depending on the set of data used for their determination. Finally, the description of the J/$\psi$ production in the NRQCD framework was recently extended to the low-$p_{\rm T}$ region, and the polarization parameters were studied in a color glass condensate (CGC) + NRQCD formalism, obtaining a fair agreement with LHC data at forward rapidity~\cite{Ma:2018qvc}. Measurements of the polarization parameters are also available for several bottomonium states, and in particular for the $\Upsilon(1{\rm S})$, $\Upsilon(2{\rm S})$ and $\Upsilon(3{\rm S})$ resonances, which were shown to exhibit little or no polarization at LHC energies~\cite{Chatrchyan:2012woa,Khachatryan:2016vxr,Aaij:2017egv}. Approaches similar to that adopted for charmonium, which also need to take into account the rather complex feed-down decay structure for these states, lead to a fair agreement with the experimental results~\cite{Han:2014kxa}.

In this Letter, we move a step forward by presenting the first measurement of J/$\psi$ and $\Upsilon(1{\rm S})$ polarization in ultrarelativistic heavy-ion interactions performed by the ALICE Collaboration by studying Pb--Pb collisions at $\sqrt{s_{\rm NN}}=5.02$ TeV. Such collisions represent an important source of information for the investigation of the phase diagram of QCD~\cite{BraunMunzinger:2008tz}, and in particular for the study of the properties of the quark--gluon plasma (QGP), a state of matter where quarks and gluons are not confined inside hadrons~\cite{Braun-Munzinger:2015hba}. Among the experimental observables studied in heavy-ion collisions the suppression of heavy quarkonium production is a fundamental signal, since QGP formation prevents the binding of the heavy-quark pair due to the screening of the color charge~\cite{Matsui:1986dk} and, more generally, has strong effects on the spectral functions~\cite{Laine:2007gj}. 
At LHC energies, another mechanism, corresponding to the (re)generation of charmonium states in the QGP and/or when the system hadronizes, becomes relevant~\cite{BraunMunzinger:2000px,Thews:2000rj}, in particular at low $p_{\rm T}$, due to the large charm-quark multiplicity ($>100$ pairs in a central Pb--Pb collision).
The presence of a deconfined system may in principle affect also the polarization of quarkonium states. In Ref.~\cite{Ioffe:2003rd} the observation of a partial transverse polarization for the J/$\psi$ was predicted in case of QGP formation, due to a modification of the non-perturbative effects in the high energy-density phase. More generally, the observed prompt J/$\psi$ are known to be a mixture of direct production and decay products from higher-mass charmonium states ($\psi(2{\rm S})$, $\chi_{\rm c}$). In nuclear collisions, since suppression effects are expected to affect more strongly the less bound states, the relative contribution of direct and feed-down production would change with respect to that in pp collisions, and the overall measured polarization may be different according to the potentially different polarization of the various states~\cite{Shao:2014fca,Sirunyan:2019apc}. On the other hand, the contribution of the regeneration mechanism in the J/$\psi$ formation process by recombination of uncorrelated ${\rm c}{\rm \overline c}$ pairs is likely to give rise to unpolarized production at low $p_{\rm T}$. 
Finally, the possible presence of polarization is known to strongly affect the acceptance for J/$\psi$ detection in the dilepton decay (up to 20--30\% in ALICE~\cite{Abelev:2011md}), and its measurement is an important requisite for an unbiased evaluation of the absolute yields in nuclear collisions. A first measurement of $\Upsilon(1{\rm S})$ polarization in Pb--Pb collisions is also presented in this Letter, even if the corresponding candidate sample is smaller by a factor $\sim$30, leading to larger uncertainties. For such a state, considerations similar to those discussed for the J/$\psi$ should hold, except that the contribution of the regeneration mechanism should be negligible due to the much lower multiplicity of bottom quarks with respect to charm.

The next sections of the Letter are organized as follows. Section 2 contains a short description of the experimental apparatus and some details on the data sample used in this analysis. The analysis procedure and the evaluation of systematic uncertainties are presented in Sec.~3, while the results on the J/$\psi$ and $\Upsilon(1{\rm S})$ polarization parameters $\lambda_{\theta}$, $\lambda_{\phi}$ and $\lambda_{\theta\phi}$ are shown in Sec.~4. The conclusions are presented in Sec.~5.

\section{Experimental setup and data sample}

The measurement described in this Letter is performed with the ALICE detector~\cite{Aamodt:2008zz,Abelev:2014ffa}, whose main components are a central barrel and a forward muon spectrometer. The latter covers the pseudorapidity region $-4<\eta<-2.5$ and is used to detect muon pairs from quarkonium decays~\cite{Aamodt:2011gj}. The muon spectrometer includes a hadron absorber made of concrete, carbon and steel with a thickness of 10 interaction lengths, followed by five tracking stations (cathode-pad chambers), with the central one embedded inside a dipole magnet with a 3 T$\cdot$m field integral. Downstream of the tracking system, an iron wall filters out the remaining hadrons as well as low-momentum muons originating from pion and kaon decays, and is followed by two trigger stations (resistive plate chambers). Another forward detector, the V0~\cite{Abbas:2013taa}, composed of two scintillator arrays located at opposite sides of the interaction point (IP) and covering the pseudorapidity intervals $-3.7<\eta < -1.7$ and $2.8 <\eta< 5.1$, provides the minimum bias (MB) trigger which is given by a coincidence of signals from the two sides. Among the central barrel detectors, the two layers of the Silicon Pixel Detector (SPD), with $|\eta|<2$ and $|\eta|<1.4$ coverage, and corresponding to the inner part of the ALICE Inner Tracking System (ITS)~\cite{Aamodt:2010aa}, are used to determine the position of the interaction vertex. Finally, the Zero Degree Calorimeters (ZDC)~\cite{ALICE:2012aa}, located on either side of the IP at $\pm$ 112.5 m along the beam axis, detect spectator nucleons emitted at zero degrees with respect to the LHC beam axis and are used to reject electromagnetic Pb--Pb interactions. 

The analysis is based on events where, in addition to the MB condition, two opposite-sign tracks are detected in the triggering system of the muon spectrometer (dimuon trigger). The dimuon trigger selects tracks each having a transverse momentum above a threshold nominally set at $p_{\rm T}^{\mu} = 1$ GeV/$c$, corresponding to the value for which the single-muon trigger efficiency reaches 50\%~\cite{Bossu:2012jt}. The single-muon trigger efficiency reaches a plateau value of 98\% at $\sim 2.5$ GeV/$c$.

The events are further characterized according to their centrality, i.e., the degree of geometric overlap of the colliding nuclei. It is estimated by means of a Glauber model fit to the V0 signal amplitude
distribution~\cite{Abelev:2013qoq,ALICE-PUBLIC-2018-011}, with more central events leading to a larger signal in the V0. In this analysis, events corresponding to the most central 90\% of the inelastic Pb--Pb cross section are selected, as for these events
the MB trigger is fully efficient and the residual contamination from electromagnetic processes is negligible.

The results of the analysis are obtained using the $\sqrt{s_{\mathrm{NN}}}$ = 5.02 TeV Pb--Pb data samples collected by the ALICE experiment during the years 2015 and 2018, corresponding to an integrated luminosity $L_{\mathrm{int}}\sim$ 750 $\mu {\rm b}^{-1}$.

\section{Data analysis}

The J/$\psi$ and $\Upsilon(1{\rm S})$ candidates are formed by combining opposite-sign muons reconstructed using the tracking algorithm described in Ref.~\cite{Aamodt:2011gj}. In order to reject tracks at the edge of the spectrometer acceptance, the condition $-4 < \eta_{\mu} < -2.5$ is required. In addition, tracks must have a radial transverse position at the end of the absorber in the range $17.6 < R_{\mathrm{abs}} < 88.9$ cm. This selection is applied to remove tracks passing through the inner and denser part of the absorber, which are strongly affected by multiple scattering. For each muon candidate, a match between tracks reconstructed in the tracking system and track segments in the muon trigger system is required.

The J/$\psi$ polarization parameters $\lambda_{\theta}$, $\lambda_{\phi}$ and $\lambda_{\theta\phi}$ are studied as a function of transverse momentum in the intervals $2<p_{\rm T}<4$, $4<p_{\rm T}<6$ and $6<p_{\rm T}<10$ GeV/$c$. For each $p_{\rm T}$ interval, a two--dimensional (2D) grid of dimuon invariant-mass spectra is created, corresponding to intervals in $\cos\theta$ and $\phi$, where $\theta$ and $\phi$ are  the polar and azimuthal emission angles, respectively, of the decay products in the J/$\psi$ rest frame, with respect to the reference axis. More in detail, the 2D grid covers the fiducial region $-0.8<\cos\theta<0.8$ (17 intervals), $0.5<\phi<\pi-0.5$ rad (8 intervals, assuming a symmetric distribution around $\phi=\pi$), with the choice of the boundaries as well as the width of the intervals dictated by acceptance considerations.

The analysis is performed choosing two different reference systems for the determination of the angular variables. In the Collins-Soper (CS) frame the $z$-axis is defined as the bisector of the angle between the direction of one beam and the opposite of the direction of the other one in the rest frame of the decaying particle, allowing therefore an evaluation of the polarization parameters with respect to the direction of motion of the colliding hadrons. In the helicity (HE) reference frame the $z$-axis is given by the direction of the decaying particle in the center-of-mass frame of the collision, and therefore the polarization can be evaluated with respect to the momentum direction of the J/$\psi$ itself. The $\phi = 0$ plane is the one containing the two beams    in the J/$\psi$ rest frame.

For each dimuon invariant-mass spectrum, the J/$\psi$ raw yield is obtained by means of a binned maximum likelihood fit in the interval $2.1 < m_{\mu\mu} < 4.9$ GeV/$c^{2}$. The background continuum is parameterized with a Gaussian distribution  whose width varies linearly with the mass or, alternatively, with a fourth degree polynomial function times an exponential. The $\jpsi$ signal is modeled with a pseudo-Gaussian function or with a Crystal Ball function with asymmetric tails on both sides of the peak~\cite{ALICE-PUBLIC-2015-006}.

\begin{figure}[ht]
\includegraphics[scale=0.40]{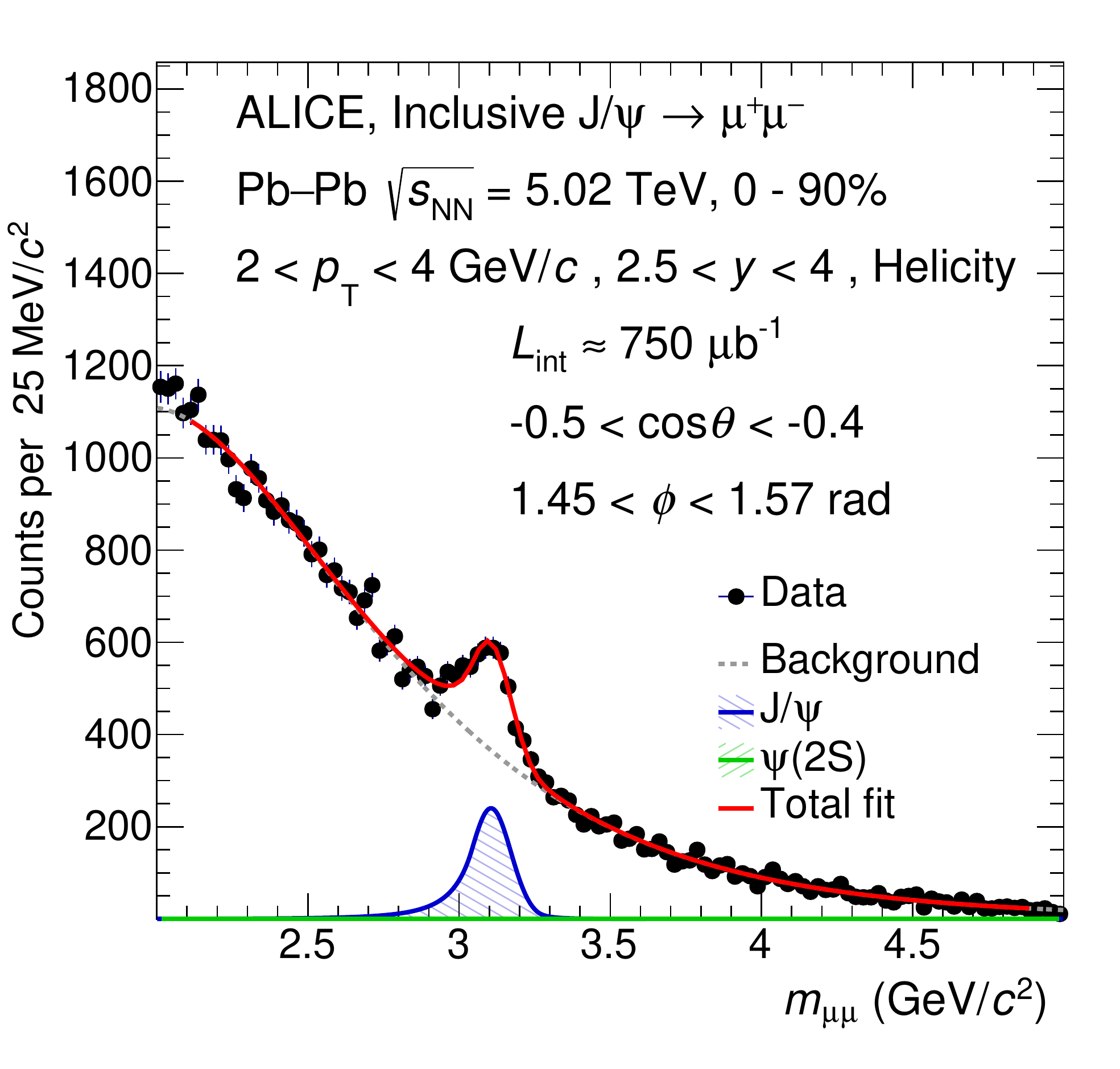}
\includegraphics[scale=0.40]{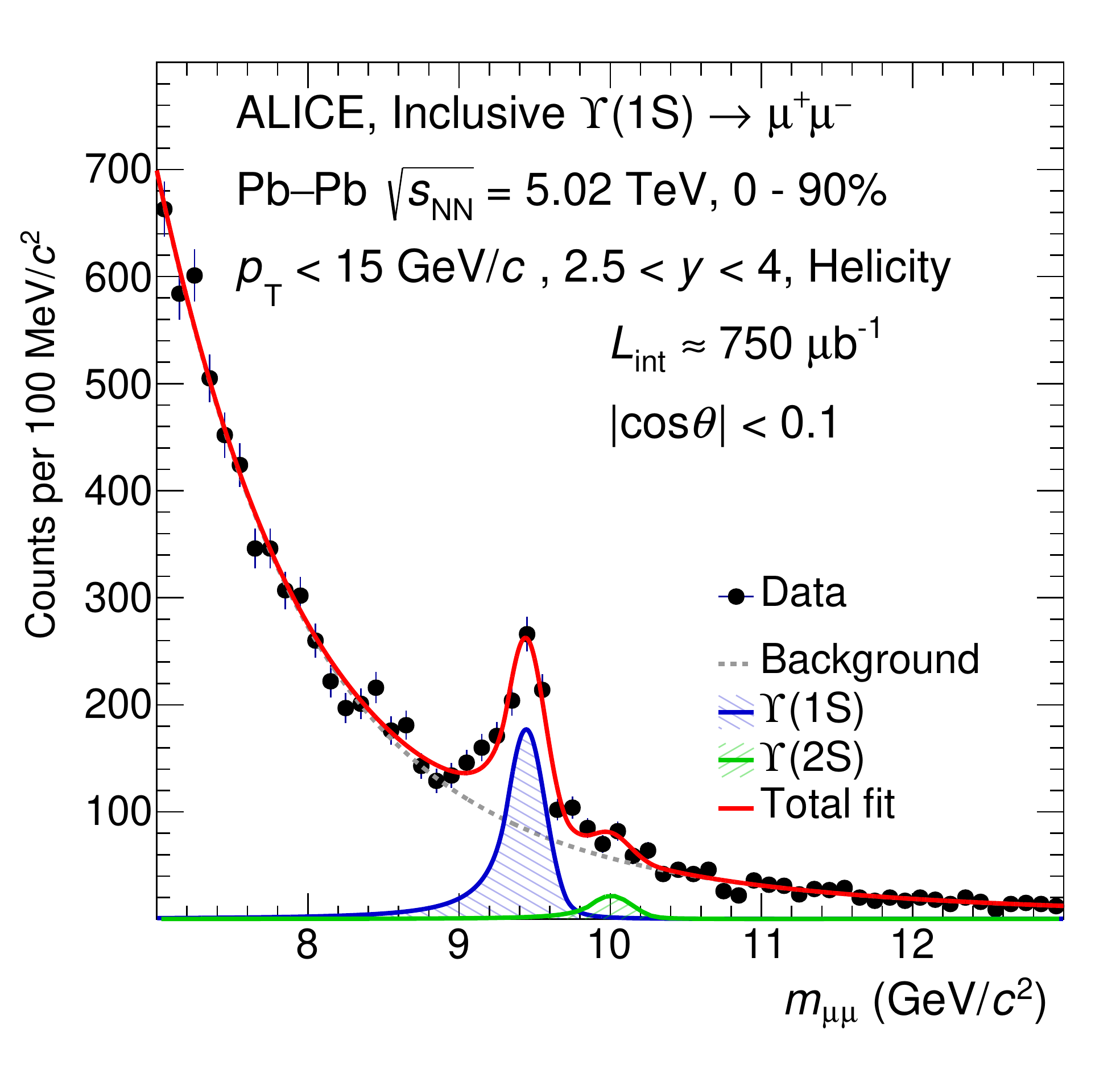}

\caption{Examples of fits to the invariant-mass distributions in the helicity reference frame. The left plot corresponds to the J/$\psi$ mass region, while  
on the right a fit to the $\Upsilon(1S)$ mass region is shown. 
The fits are performed using an extended Crystal Ball function for the resonance  signals, while the background is parameterized with a variable width Gaussian.} 
\label{fig:signal_extraction_example}
\end{figure}

The $\jpsi$ mass is kept free in the fits, while for each interval ($i$,$j$) in ($\cos\theta$,$\phi$) the width is fixed to $\sigma_{{\rm J}/\psi}^{i,j}=\sigma_{{\rm J}/\psi}^{i,j,{\rm MC}} \cdot (\sigma_{{\rm J}/\psi}/\sigma_{{\rm J}/\psi}^{\mathrm{MC}})$, i.e., scaling the resonance width extracted from  Monte Carlo (MC) simulations ($\sigma_{{\rm J}/\psi}^{i,j,\rm {MC}}$) by the ratio between the width obtained by fitting the angle-integrated spectrum in data ($\sigma_{{\rm J}/\psi}$) and MC ($\sigma_{{\rm J}/\psi}^{\mathrm{MC}}$) for the $p_{\rm T}$ interval under consideration. The parameters of the non-Gaussian tails of the resonance are kept fixed to the MC values.
The $\psi(2\mathrm{S})$ contribution, although comparatively negligible, is also taken into account in the fits, with its width and mass fixed in each fit to those of the J/$\psi$ according to the relations $\sigma_{\psi(2\mathrm{S})}=\sigma_{{\rm J}/\psi} \cdot \sigma_{\psi(2\mathrm{S})}^{\mathrm{MC}}/\sigma_{{\rm J}/\psi}^{\mathrm{MC}}$ and $m_{\psi(2\mathrm{S})}=m_{{\rm J}/\psi} + m_{\psi(2\mathrm{S})}^{\mathrm{PDG}} - m_{{\rm J}/\psi}^{\mathrm{PDG}}$, with the Particle Data Group (PDG) masses taken from Ref.~\cite{Tanabashi:2018oca}. 
In Fig.~\ref{fig:signal_extraction_example} (left)  an example of a fit to the invariant-mass spectrum in the J/$\psi$ mass region is shown. Due to the stability of the extracted J/$\psi$ parameters (mass, width), the fits were carried out directly on the sum of the 2015 and 2018 invariant mass spectra.

The $\jpsi$ raw yields as a function of the angular variables are then corrected by the product of the acceptance and detector efficiency ($A\times\varepsilon$), which is evaluated as a function of $\cos\theta$ and $\phi$ on a 2D grid via MC simulations. The $\jpsi$ are  generated according to $\pt$ and $y$ distributions directly tuned on data~\cite{Acharya:2019iur} via an iterative procedure~\cite{Acharya:2018kxc}, and their decay muons are propagated inside a realistic description of the ALICE setup, based on GEANT 3.21 \cite{Brun:1082634}. The misalignment of the detection elements and the time-dependent status of each electronic channel during the data taking period are taken into account as well.
In the $\jpsi$ generation an isotropic distribution of decay products, corresponding to the assumption of no polarization, is adopted. 
Due  to the choice of relatively small ($\cos\theta$, $\phi$) intervals, the $A\times\varepsilon$ values for each interval are quite insensitive to the specific angular distribution assumed in the generation.

\begin{figure}[ht]
	\includegraphics[scale=0.40]{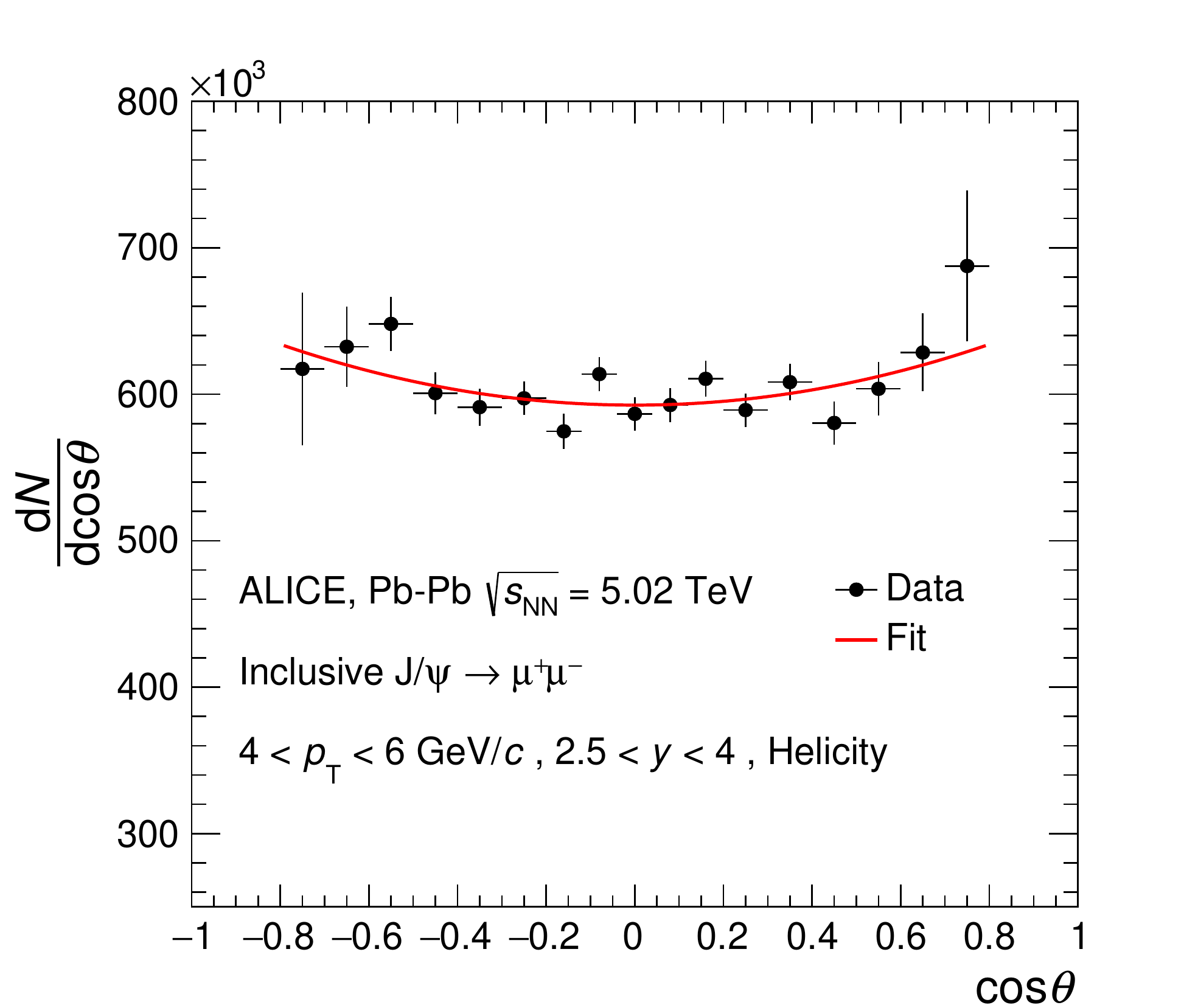}
	\includegraphics[scale=0.40]{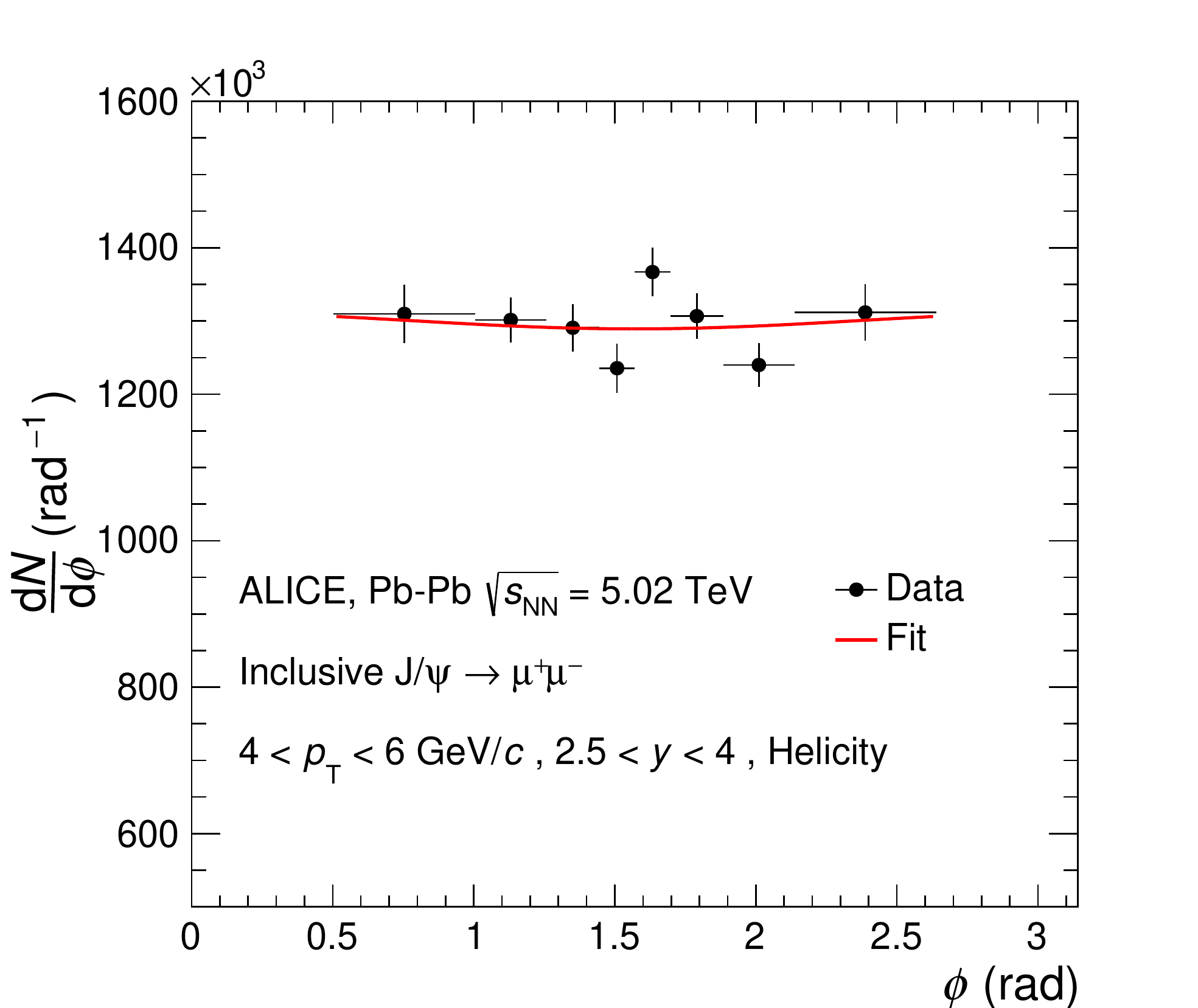}
	\caption{\label{polarization_fit} Fit to the $\jpsi$ 2D angular distributions in the helicity reference frame projected along $\cos\theta$ (left) and $\phi$ (right) for $2.5<y<4$ and $4<\pt<6$ GeV/$c$. The displayed uncertainties are statistical.}
\end{figure}

The three polarization parameters $\lambda_{\theta}$, $\lambda_{\phi}$ and $\lambda_{\theta\phi}$ are obtained through $\chi^2$-minimization fits of the 2D  $\jpsi$ distributions, corrected for acceptance and efficiency,  according to Eq.~\ref{eq:1}. For each combination of signal and background shape used in the fit to the dimuon invariant-mass spectra, a separate evaluation of the polarization parameters is carried out and their average is taken as the best estimate. The statistical uncertainty is given by the average of the statistical uncertainties of the 2D fits, while the root mean square of the results provides the systematic uncertainty on the signal extraction, with the absolute values ranging between 0.002 and 0.039. The overall procedure described above was checked beforehand with a MC closure test. The 2D fits on the ($\cos\theta$, $\phi$) distributions only allow a determination of the absolute value of $\lambdaThetaPhi$, due to the presence of $\sin 2\theta$ in the corresponding term that induces an ambiguity in its sign. It is checked that the values of $\lambdaTheta$ and $\lambdaPhi$ are stable against the choice of the sign of the $\lambdaThetaPhi$ term. In the following the $\lambdaThetaPhi$ values corresponding to the choice of a positive sign are quoted.
Figure~\ref{polarization_fit} illustrates an example of the fit to the angular distributions. For better visibility, both the distribution and the fitted function are projected along one dimension. 

In addition to the systematic uncertainty related to the choice of the mass shapes for signal and background, several other sources are taken into account. 
First, an alternative procedure for extracting the $\jpsi$ signal is carried out, by keeping its width as a free parameter in the invariant-mass fits. The corresponding results for the polarization parameters are then obtained and the averages of the values corresponding to fixing the width or not are taken as the central values for $\lambdaTheta$, $\lambdaPhi$ and $\lambdaThetaPhi$. Half  the difference between the results obtained with free or MC-anchored widths is then considered as a further systematic uncertainty related to the signal extraction. This uncertainty is found to be the leading contribution to the total absolute systematic uncertainty on the polarization parameters, and ranges between 0.001 and 0.063, the latter value corresponding to the uncertainty on $\lambda_{\theta}^{\rm HE}$ for $2<p_{\rm T}<4$ GeV/$c$.

Another source of systematic uncertainty is related to the evaluation of the trigger efficiency. The muon trigger response function as a function of the single muon transverse momentum $p_{\rm T}^\mu$ can be obtained via MC or with a procedure based on data~\cite{Abelev:2014qha}. Small deviations are found for $\ensuremath{p_{\mathrm{\textsc{t}}}}^{\mu} < 2$ GeV/$c$ which induce an effect on $A\times\varepsilon$ for the $\jpsi$. Therefore, the polarization parameters are re-calculated with $A\times\varepsilon$ values weighted in such a way to account for the deviations. The variation of the polarization parameters between the different trigger efficiency estimates is taken as the related systematic uncertainty, with values ranging from 0.001 to 0.043, the highest values being found for $\lambda_{\theta}^{\rm HE}$ in $2<p_{\rm T}<4$ GeV/$c$. The systematic uncertainty related to the evaluation of the muon tracking efficiency is found to be negligible for this analysis, allowing a significant reduction of the total systematic uncertainty with respect to  previous pp analyses~\cite{Acharya:2018uww}. Indeed, although the difference between efficiencies calculated via MC or from data~\cite{Abelev:2014qha} is of the order of 2\%, a detailed investigation has shown no dependence on the angular variables and therefore no effect on the polarization parameters.

Finally, the systematic uncertainty induced by the choice of the \pt and $y$ distributions used as an input for the calculation of $A\times\varepsilon$ is evaluated testing alternative $\pt$ and $y$ parameterizations, which are obtained by varying within their uncertainties the default distributions directly tuned on Pb--Pb data. The polarization parameters extracted with the modified values of  $A\times\varepsilon$ are compared with those obtained with the default input shapes and the corresponding systematic uncertainty extracted in this way is found to range between 0.001 and 0.030, with the largest value assigned to $\lambda_{\theta}^{\rm HE}$ for $2<p_{\rm T}<4$ GeV/$c$. The influence of the choice of the angular distributions of the J/$\psi$ decay products for the $A\times\varepsilon$ calculation is also investigated by means of an iterative procedure on these input distributions. The effect is found to be negligible, also due to the fact that the 2D correction procedure on the angular variables is by definition relatively insensitive to the specific choice of the corresponding distributions. A summary of the values of all the absolute systematic uncertainties, which are considered as uncorrelated as a function of $p_{\rm T}$, is reported in Table~\ref{table:jpsi_systematic_uncertainties_summary}. The total systematic uncertainties are obtained, for each parameter and $p_{\rm T}$ interval, as the quadratic sum of the values.

\begin{table}[htp]
		\caption{Summary of the absolute systematic uncertainties on the evaluation of the  $\jpsi$ polarization parameters. All the uncertainties are considered as uncorrelated as a function of $p_{\rm T}$.}\label{table:jpsi_systematic_uncertainties_summary}
	\begin{center}
		\begin{tabular}{cccccc|cccc}
			\hline
			& & \multicolumn{4}{c}{Helicity} & \multicolumn{4}{c}{Collins-Soper} \\
			\cline{3-10}
			&  & Signal & J/$\psi$ & Trigger & Input & Signal & J/$\psi$ & Trigger & Input \\
			& $\pt$ (GeV/\textit{c}) & extr. & width & eff. & MC & extr. & width & eff. & MC \\
			\cline{3-10}
			\multirow{3}{*}{\lambdaTheta} 
			& 2$<$\pt$<$4  & 0.030 & 0.063 & 0.043 & 0.030 & 0.026 & 0.049 & 0.015 & 0.009 \\ 
			& 4$<$\pt$<$6  & 0.017 & 0.046 & 0.040 & 0.024 & 0.002 & 0.052 & 0.018 & 0.007 \\ 
			& 6$<$\pt$<$10 & 0.039 & 0.005 & 0.018 & 0.017 & 0.022 & 0.001 & 0.011 & 0.006 \\
			\cline{3-10}
			\multirow{3}{*}{\lambdaPhi} 
			& 2$<$\pt$<$4  & 0.007 & 0.030 & 0.004 & 0.002 & 0.024 & 0.010 & 0.020 & 0.003 \\ 
			& 4$<$\pt$<$6  & 0.003 & 0.035 & 0.003 & 0.003 & 0.002 & 0.010 & 0.020 & 0.003 \\ 
			& 6$<$\pt$<$10 & 0.002 & 0.009 & 0.001 & 0.002 & 0.005 & 0.013 & 0.011 & 0.002 \\
			\cline{3-10}
			\multirow{3}{*}{\lambdaThetaPhi} 
			& 2$<$\pt$<$4  & 0.021 & 0.029 & 0.024 & 0.001 & 0.013 & 0.010 & 0.017 & 0.015 \\ 
			& 4$<$\pt$<$6  & 0.007 & 0.011 & 0.017 & 0.006 & 0.002 & 0.042 & 0.010 & 0.015 \\ 
			& 6$<$\pt$<$10 & 0.020 & 0.019 & 0.007 & 0.008 & 0.007 & 0.042 & 0.003 & 0.013 \\
			\hline
		\end{tabular}

	\end{center}
\end{table}

A similar procedure is followed for the extraction of the $\Upsilon(1{\rm S})$ polarization parameters. Due to the smaller candidate sample, integrated values over the kinematic interval $2.5<y<4$, $p_{\rm T}<15$ GeV/$c$ are obtained. The main difference with respect to the 2D approach followed for the J/$\psi$ is the use of a simultaneous fit to the 1D angular distributions~\cite{Acharya:2018uww}, after integration over the other variables. The requirement $p_{\rm T}^{\mu}>2$ GeV/$c$, which helps reducing the combinatorial background, is included~\cite{Acharya:2018mni}. The $\Upsilon(1{\rm S})$ signal extraction in the various $\cos\theta$ and $\phi$ intervals is performed by means of invariant-mass fits (see the right panel of  Fig.~\ref{fig:signal_extraction_example} for an example). The functions chosen for the resonances are the same as in the J/$\psi$ analysis (pseudo-Gaussian or Crystal Ball), the mass value is fixed to that obtained from a fit to the integrated invariant-mass distribution, while the width for each angular interval is fixed to the MC value scaled by the ratio of the widths between data and MC for the angle-integrated distributions. The tail parameters are fixed to MC values. The small contribution from $\Upsilon(2{\rm S})$ is also included in the fits~\cite{Acharya:2018mni}. The background continuum is parameterized with a Gaussian distribution  whose width varies linearly with the mass or, alternatively, with a second degree polynomial function times an exponential. The systematic uncertainty on the signal extraction is calculated with the same procedure adopted for the J/$\psi$.  An uncertainty related to the choice of the signal width has also been considered, taken as the half-difference between the results obtained with the prescription described above and using as an alternative prescription the pure MC values. The uncertainty on the trigger efficiency is negligible, due to the additional requirement on the single-muon transverse momentum which selects a $p_{\rm T}$-region where the trigger efficiency is very high and its evaluation via data and MC is consistent. Finally, the procedure for the determination of the uncertainty related to the $\Upsilon(1{\rm S})$ kinematic distributions used in the MC is the same as for the J/$\psi$. The total systematic uncertainties for the $\Upsilon(1{\rm S})$ analysis are reported in Table~\ref{table:UpsilonResults}, together with the results.
\section{Results}

\begin{figure}[h]
\centering
\includegraphics[width=1.0\linewidth]{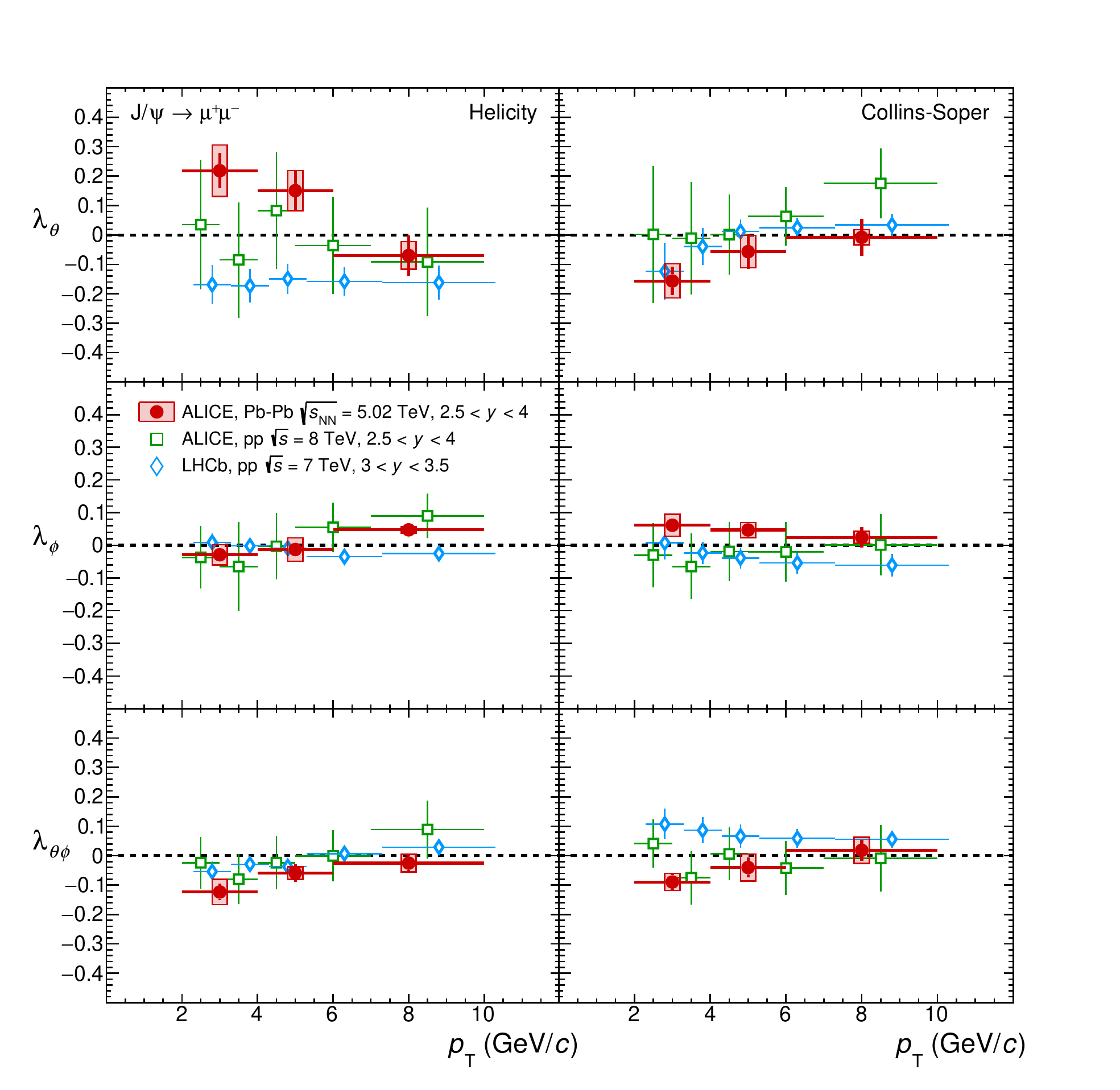}
\caption{Inclusive $\jpsi$ polarization parameters as a function of transverse momentum for Pb--Pb collisions at $\ensuremath{\sqrt{s_{_{\mathrm{NN}}}}}=$ 5.02 TeV, compared with results obtained in pp collisions by ALICE at $\ensuremath{\sqrt{s}}=$ 8 TeV \cite{Acharya:2018uww} and by LHCb for prompt $\jpsi$ at $\ensuremath{\sqrt{s}}=$ 7 TeV \cite{Aaij:2013nlm} (the LHCb markers were  shifted horizontally by +0.3 GeV/\textit{c} for better visibility) in the rapidity interval $3<y<3.5$. The error bars represent the total uncertainties for the pp results, while for Pb--Pb statistical and systematic uncertainties are plotted separately as a vertical bar and a shaded box, respectively. In the left part of the plot the polarization parameters in the helicity reference frame are reported, in the right those for the Collins-Soper frame.}
\label{fig:polarization_parameters}
\end{figure}

The polarization parameters for $\jpsi$ inclusive production in Pb--Pb collisions at $\ensuremath{\sqrt{s_{_{\mathrm{NN}}}}}=$ 5.02 TeV in the helicity and Collins-Soper reference frames are shown in Fig.~\ref{fig:polarization_parameters} and the corresponding numerical values are reported in Table~ \ref{table:polarization_parameters}. In Fig. \ref{fig:polarization_parameters}, $\lambda_{\theta}$, $\lambda_{\phi}$ and $\lambda_{\theta\phi}$ are also compared with the LHCb \cite{Aaij:2013nlm} and ALICE \cite{Acharya:2018uww} measurements in pp collisions at $\ensuremath{\sqrt{s}}=$ 7 and 8 TeV, respectively. 

\begin{table}[htp]
		\caption{$\jpsi$ polarization parameters, measured for Pb--Pb collisions at $\sqrt{s_{\rm NN}}=5.02$ TeV, in the helicity and Collins-Soper reference frames in the rapidity interval $2.5<y<4$. The first uncertainty is statistical and the second systematic.}\label{table:polarization_parameters}
	\begin{center}
		\begin{tabular}{cccc}
			& $\pt$ (GeV/\textit{c}) & Helicity & Collins-Soper \\
			\hline
			\multirow{3}{*}{$\lambdaTheta$} 
			& 2$<$\pt$<$4  & $0.218 \pm 0.060 \pm 0.087$ & $-0.157 \pm 0.049 \pm 0.058$ \\ 
            & 4$<$\pt$<$6  & $0.151 \pm 0.071 \pm 0.068$ & $-0.057 \pm 0.059 \pm 0.055$ \\ 
            & 6$<$\pt$<$10 & $-0.070 \pm 0.068 \pm 0.047$ & $-0.008 \pm 0.063 \pm 0.026$ \\ 
            \hline 
            \multirow{3}{*}{$\lambdaPhi$} 
            & 2$<$\pt$<$4  & $-0.029 \pm 0.017 \pm 0.031$ & $0.061 \pm 0.015 \pm 0.033$ \\ 
            & 4$<$\pt$<$6  & $-0.013 \pm 0.019 \pm 0.036$ & $0.047 \pm 0.024 \pm 0.023$ \\ 
            & 6$<$\pt$<$10 & $0.047 \pm 0.021 \pm 0.010$ & $0.024 \pm 0.032 \pm 0.018$ \\ 
            \hline 
            \multirow{3}{*}{$\lambdaThetaPhi$}
            & 2$<$\pt$<$4  & $-0.124 \pm 0.028 \pm 0.043$ & $-0.090 \pm 0.027 \pm 0.029$ \\ 
            & 4$<$\pt$<$6  & $-0.059 \pm 0.030 \pm 0.021$ & $-0.040 \pm 0.034 \pm 0.046$ \\ 
            & 6$<$\pt$<$10 & $-0.025 \pm 0.031 \pm 0.030$ & $0.018 \pm 0.035 \pm 0.044$ \\
            \hline
		\end{tabular}
	\end{center}
\end{table} 

For all the \pt intervals and in both reference frames the values of the polarization parameters exhibit at most slight deviations from zero. In particular, $\lambda_\theta^{\rm HE}$ indicates a slight transverse polarization at low $p_{\rm T}$ ($\sim 2.1\sigma$ effect, calculated using the Gaussian approximation), while $\lambda_\theta^{\rm CS}$ shows a weak longitudinal polarization ($\sim  2.1\sigma$). When increasing $p_{\rm T}$, the central values of $\lambda_\theta$ become close to zero.  All values of $\lambda_{\phi}$ and $\lambda_{\theta\phi}$ are, in absolute value, smaller than 0.1, except for $\lambda_{\theta\phi}^{\rm HE}$, which is $-0.124$ at low $p_{\rm T}$ and deviates from zero by $\sim 2.4\sigma$.

When comparing with the pp results, no significant difference is found with respect to ALICE results at $\sqrt{s}= 8$ TeV, which are compatible with zero. A significant difference is found with respect to the  higher-precision LHCb results at $\sqrt{s}= 7$ TeV, reaching 3.3$\sigma$ in the interval $2<p_{\rm T}<4$ GeV/$c$  in the helicity reference frame, where pp data~\cite{Aaij:2013nlm} indicate a small but significant degree of longitudinal polarization, while the Pb--Pb results favor a slightly transverse polarization. In Pb--Pb collisions at LHC energies, a significant fraction of the detected J/$\psi$ originates from the recombination of ${\rm c}{\rm\overline{c}}$ pairs in the QGP phase or when the system hadronizes. Moreover, the contribution from higher-mass charmonium states decaying to J/$\psi$ could vary between pp and Pb-Pb due to different suppression effects for each state in nuclear collisions. Therefore, the observed hint for a different polarization in pp and Pb--Pb might be a reflection of the different production and suppression mechanisms in the two systems, but more precise data, along with quantitative theory estimates, are needed for a definite conclusion. It should also be noted that the ALICE results refer to inclusive production, while LHCb has measured prompt J/$\psi$. However, as discussed in Ref.~\cite{Abelev:2011md}, the size of the non-prompt component is small in the covered $p_{\rm T}$ region (of the order of 15\% at high $p_{\rm T}$) and its polarization was also measured to be small by CDF ($\lambda_{\theta}^{\rm{HE}}\sim -0.1$~\cite{Abulencia:2007us}), implying that the net effect of this source on inclusive J/$\psi$ polarization should be negligible. 

In Table~\ref{table:UpsilonResults} the values of the $\Upsilon(1{\rm S})$ polarization parameters are shown. The $\lambda_{\theta}$ values are consistent with zero, with large uncertainties that prevent a firm conclusion on the absence of polarization in nuclear collisions. The $\lambdaPhi$ and $\lambdaThetaPhi$ values are also consistent with zero. The relatively smaller uncertainties for these parameters are related to a more uniform acceptance distribution as a function of the azimuthal angular variable.

\begin{table}[htbp]
		\caption{$\Upsilon(1{\rm S})$ polarization parameters in the helicity and Collins-Soper reference frames measured in Pb--Pb collisions at $\sqrt{s_{\rm NN}}=5.02$ TeV in the rapidity interval $2.5<y<4$ and for transverse momentum $p_{\rm T}<15$ GeV/$c$. The first uncertainty is statistical and the second systematic.}
	\begin{center}
		\begin{tabular}{ccc}
		
			& {Helicity} & {Collins-Soper} \\
			\hline
			{$\lambdaTheta$} & {$-0.090 \pm 0.395 \pm 0.101$} & {$0.418 \pm 0.526 \pm 0.178$} \\ 
			\hline
			{$\lambdaPhi$} & {$-0.094 \pm 0.072 \pm 0.020$} & {$-0.141 \pm 0.087 \pm 0.033$} \\ 
			\hline
			{$\lambdaThetaPhi$} & {$-0.074 \pm 0.099 \pm 0.020$} & {$0.017 \pm 0.113 \pm 0.024$}\\ 
			\hline
			
		\end{tabular}
		\label{table:UpsilonResults}
	\end{center}
\end{table}

\section{Conclusions}

The first measurement of the polarization parameters for J/$\psi$ production in nuclear collisions at LHC energies was carried out by the ALICE Collaboration  in Pb--Pb interactions at $\sqrt{s_{\rm NN}}=5.02$ TeV. The $\lambda_{\theta}$, $\lambda_{\phi}$ and  $\lambda_{\theta\phi}$ parameters were evaluated in the helicity and Collins-Soper reference frames in the rapidity interval $2.5<y<4$ and in the transverse momentum interval $2<p_{\rm T}<10$ GeV/$c$. All the parameter values are close to zero, with a $\sim 2.1\sigma$ indication for a small transverse polarization in the helicity frame at low $p_{\rm T}$, and a corresponding indication for a small longitudinal polarization in the Collins-Soper frame ($\sim 2.1\sigma$ effect). When comparing these results with pp data taken at higher energy at the LHC, an interesting feature is a significant difference in $\lambda_{\theta}^{\rm HE}$ with respect to the LHCb results which showed  instead a small longitudinal polarization in a similar kinematic domain. This first result obtained for J/$\psi$ in nuclear collisions and described in this Letter represents therefore a starting point for future studies connecting such features with the known differences in the production mechanisms between pp and nucleus--nucleus collisions. Results were also obtained for the first time for the $\Upsilon(1{\rm S})$ polarization, integrated over $p_{\rm T}$ and $y$, showing, within the large uncertainties of the measurement, values compatible with the absence of polarization.


\newenvironment{acknowledgement}{\relax}{\relax}
\begin{acknowledgement}
\section*{Acknowledgements}

The ALICE Collaboration would like to thank all its engineers and technicians for their invaluable contributions to the construction of the experiment and the CERN accelerator teams for the outstanding performance of the LHC complex.
The ALICE Collaboration gratefully acknowledges the resources and support provided by all Grid centres and the Worldwide LHC Computing Grid (WLCG) collaboration.
The ALICE Collaboration acknowledges the following funding agencies for their support in building and running the ALICE detector:
A. I. Alikhanyan National Science Laboratory (Yerevan Physics Institute) Foundation (ANSL), State Committee of Science and World Federation of Scientists (WFS), Armenia;
Austrian Academy of Sciences, Austrian Science Fund (FWF): [M 2467-N36] and Nationalstiftung f\"{u}r Forschung, Technologie und Entwicklung, Austria;
Ministry of Communications and High Technologies, National Nuclear Research Center, Azerbaijan;
Conselho Nacional de Desenvolvimento Cient\'{\i}fico e Tecnol\'{o}gico (CNPq), Financiadora de Estudos e Projetos (Finep), Funda\c{c}\~{a}o de Amparo \`{a} Pesquisa do Estado de S\~{a}o Paulo (FAPESP) and Universidade Federal do Rio Grande do Sul (UFRGS), Brazil;
Ministry of Education of China (MOEC) , Ministry of Science \& Technology of China (MSTC) and National Natural Science Foundation of China (NSFC), China;
Ministry of Science and Education and Croatian Science Foundation, Croatia;
Centro de Aplicaciones Tecnol\'{o}gicas y Desarrollo Nuclear (CEADEN), Cubaenerg\'{\i}a, Cuba;
Ministry of Education, Youth and Sports of the Czech Republic, Czech Republic;
The Danish Council for Independent Research | Natural Sciences, the VILLUM FONDEN and Danish National Research Foundation (DNRF), Denmark;
Helsinki Institute of Physics (HIP), Finland;
Commissariat \`{a} l'Energie Atomique (CEA) and Institut National de Physique Nucl\'{e}aire et de Physique des Particules (IN2P3) and Centre National de la Recherche Scientifique (CNRS), France;
Bundesministerium f\"{u}r Bildung und Forschung (BMBF) and GSI Helmholtzzentrum f\"{u}r Schwerionenforschung GmbH, Germany;
General Secretariat for Research and Technology, Ministry of Education, Research and Religions, Greece;
National Research, Development and Innovation Office, Hungary;
Department of Atomic Energy Government of India (DAE), Department of Science and Technology, Government of India (DST), University Grants Commission, Government of India (UGC) and Council of Scientific and Industrial Research (CSIR), India;
Indonesian Institute of Science, Indonesia;
Centro Fermi - Museo Storico della Fisica e Centro Studi e Ricerche Enrico Fermi and Istituto Nazionale di Fisica Nucleare (INFN), Italy;
Institute for Innovative Science and Technology , Nagasaki Institute of Applied Science (IIST), Japanese Ministry of Education, Culture, Sports, Science and Technology (MEXT) and Japan Society for the Promotion of Science (JSPS) KAKENHI, Japan;
Consejo Nacional de Ciencia (CONACYT) y Tecnolog\'{i}a, through Fondo de Cooperaci\'{o}n Internacional en Ciencia y Tecnolog\'{i}a (FONCICYT) and Direcci\'{o}n General de Asuntos del Personal Academico (DGAPA), Mexico;
Nederlandse Organisatie voor Wetenschappelijk Onderzoek (NWO), Netherlands;
The Research Council of Norway, Norway;
Commission on Science and Technology for Sustainable Development in the South (COMSATS), Pakistan;
Pontificia Universidad Cat\'{o}lica del Per\'{u}, Peru;
Ministry of Science and Higher Education, National Science Centre and WUT ID-UB, Poland;
Korea Institute of Science and Technology Information and National Research Foundation of Korea (NRF), Republic of Korea;
Ministry of Education and Scientific Research, Institute of Atomic Physics and Ministry of Research and Innovation and Institute of Atomic Physics, Romania;
Joint Institute for Nuclear Research (JINR), Ministry of Education and Science of the Russian Federation, National Research Centre Kurchatov Institute, Russian Science Foundation and Russian Foundation for Basic Research, Russia;
Ministry of Education, Science, Research and Sport of the Slovak Republic, Slovakia;
National Research Foundation of South Africa, South Africa;
Swedish Research Council (VR) and Knut \& Alice Wallenberg Foundation (KAW), Sweden;
European Organization for Nuclear Research, Switzerland;
Suranaree University of Technology (SUT), National Science and Technology Development Agency (NSDTA) and Office of the Higher Education Commission under NRU project of Thailand, Thailand;
Turkish Atomic Energy Agency (TAEK), Turkey;
National Academy of  Sciences of Ukraine, Ukraine;
Science and Technology Facilities Council (STFC), United Kingdom;
National Science Foundation of the United States of America (NSF) and United States Department of Energy, Office of Nuclear Physics (DOE NP), United States of America.    
\end{acknowledgement}

\bibliographystyle{utphys}  
\bibliography{draft.bib}

\providecommand{\href}[2]{#2}\begingroup\raggedright\begin{thebibliography}{10}

\bibitem{Tanabashi:2018oca}
{\bfseries Particle Data Group} Collaboration, M.~Tanabashi {\em et~al.},
  ``{Review of Particle Physics},''
\href{http://dx.doi.org/10.1103/PhysRevD.98.030001}{{\em Phys. Rev.} {\bfseries
  D98} no.~3, (2018) 030001}.

\bibitem{Bodwin:1994jh}
G.~T. Bodwin, E.~Braaten, and G.~P. Lepage, ``{Rigorous QCD analysis of
  inclusive annihilation and production of heavy quarkonium},''
  \href{http://dx.doi.org/10.1103/PhysRevD.55.5853,
  10.1103/PhysRevD.51.1125}{{\em Phys. Rev.} {\bfseries D51} (1995)
  1125--1171}, \href{http://arxiv.org/abs/hep-ph/9407339}{{\ttfamily
  arXiv:hep-ph/9407339 [hep-ph]}}.
[Erratum: Phys. Rev.D55,5853(1997)].

\bibitem{Acharya:2017hjh}
{\bfseries ALICE} Collaboration, S.~Acharya {\em et~al.}, ``{Energy dependence
  of forward-rapidity $\mathrm {J}/\psi $ and $\psi \mathrm {(2S)}$ production
  in pp collisions at the LHC},''
  \href{http://dx.doi.org/10.1140/epjc/s10052-017-4940-4}{{\em Eur. Phys. J.}
  {\bfseries C77} no.~6, (2017) 392},
\href{http://arxiv.org/abs/1702.00557}{{\ttfamily arXiv:1702.00557 [hep-ex]}}.

\bibitem{Acharya:2019lkw}
{\bfseries ALICE} Collaboration, S.~Acharya {\em et~al.}, ``{Inclusive J/$\psi$
  production at mid-rapidity in pp collisions at $ \sqrt{s} $ = 5.02 TeV},''
  \href{http://dx.doi.org/10.1007/JHEP10(2019)084}{{\em JHEP} {\bfseries 10}
  (2019) 084}, \href{http://arxiv.org/abs/1905.07211}{{\ttfamily
  arXiv:1905.07211 [nucl-ex]}}.

\bibitem{Aaij:2015rla}
{\bfseries LHCb} Collaboration, R.~Aaij {\em et~al.}, ``{Measurement of forward
  $J/\psi$ production cross-sections in $pp$ collisions at $\sqrt{s}=13$
  TeV},'' \href{http://dx.doi.org/10.1007/JHEP05(2017)063,
  10.1007/JHEP10(2015)172}{{\em JHEP} {\bfseries 10} (2015) 172},
  \href{http://arxiv.org/abs/1509.00771}{{\ttfamily arXiv:1509.00771
  [hep-ex]}}.
[Erratum: JHEP05,063(2017)].

\bibitem{Sirunyan:2017qdw}
{\bfseries CMS} Collaboration, A.~M. Sirunyan {\em et~al.}, ``{Measurement of
  quarkonium production cross sections in pp collisions at $\sqrt{s}=$ 13
  TeV},'' \href{http://dx.doi.org/10.1016/j.physletb.2018.02.033}{{\em Phys.
  Lett.} {\bfseries B780} (2018) 251--272},
\href{http://arxiv.org/abs/1710.11002}{{\ttfamily arXiv:1710.11002 [hep-ex]}}.

\bibitem{Aad:2015duc}
{\bfseries ATLAS} Collaboration, G.~Aad {\em et~al.}, ``{Measurement of the
  differential cross-sections of prompt and non-prompt production of $J/\psi $
  and $\psi (2\mathrm {S})$ in $pp$ collisions at $\sqrt{s} = 7$ and 8 TeV with
  the ATLAS detector},''
  \href{http://dx.doi.org/10.1140/epjc/s10052-016-4050-8}{{\em Eur. Phys. J.}
  {\bfseries C76} no.~5, (2016) 283},
\href{http://arxiv.org/abs/1512.03657}{{\ttfamily arXiv:1512.03657 [hep-ex]}}.

\bibitem{Acharya:2019iok}
{\bfseries PHENIX} Collaboration, U.~Acharya {\em et~al.}, ``{$J/\psi$ and
  $\psi(2S)$ production at forward rapidity in $p$+$p$ collisions at
  $\sqrt{s}=510$ GeV},''
  \href{http://dx.doi.org/10.1103/PhysRevD.101.052006}{{\em Phys. Rev. D}
  {\bfseries 101} no.~5, (2020) 052006},
  \href{http://arxiv.org/abs/1912.13424}{{\ttfamily arXiv:1912.13424
  [hep-ex]}}.

\bibitem{Adam:2019mrg}
{\bfseries STAR} Collaboration, J.~Adam {\em et~al.}, ``{Measurements of the
  transverse-momentum-dependent cross sections of $J/\psi$ production at
  mid-rapidity in proton+proton collisions at $\sqrt{s} =$ 510 and 500 GeV with
  the STAR detector},''
  \href{http://dx.doi.org/10.1103/PhysRevD.100.052009}{{\em Phys. Rev.}
  {\bfseries D100} no.~5, (2019) 052009},
\href{http://arxiv.org/abs/1905.06075}{{\ttfamily arXiv:1905.06075 [hep-ex]}}.

\bibitem{Abe:1997jz}
{\bfseries CDF} Collaboration, F.~Abe {\em et~al.}, ``{$J/\psi$ and $\psi(2S)$
  production in $p\bar{p}$ collisions at $\sqrt{s} = 1.8$ TeV},''
\href{http://dx.doi.org/10.1103/PhysRevLett.79.572}{{\em Phys. Rev. Lett.}
  {\bfseries 79} (1997) 572--577}.

\bibitem{Abachi:1996jq}
{\bfseries D0} Collaboration, S.~Abachi {\em et~al.}, ``{$J/\psi$ production in
  $p \bar{p}$ collisions at $\sqrt{s}$ = 1.8-TeV},''
\href{http://dx.doi.org/10.1016/0370-2693(96)00067-6}{{\em Phys. Lett.}
  {\bfseries B370} (1996) 239--248}.

\bibitem{Aaij:2018pfp}
{\bfseries LHCb} Collaboration, R.~Aaij {\em et~al.}, ``{Measurement of
  $\Upsilon$ production in $pp$ collisions at $\sqrt{s}$= 13 TeV},''
  \href{http://dx.doi.org/10.1007/JHEP07(2018)134,
  10.1007/JHEP05(2019)076}{{\em JHEP} {\bfseries 07} (2018) 134},
  \href{http://arxiv.org/abs/1804.09214}{{\ttfamily arXiv:1804.09214
  [hep-ex]}}.
[Erratum: JHEP05,076(2019)].

\bibitem{Adam:2015rta}
{\bfseries ALICE} Collaboration, J.~Adam {\em et~al.}, ``{Inclusive quarkonium
  production at forward rapidity in pp collisions at $\sqrt{s}=8$ TeV},''
  \href{http://dx.doi.org/10.1140/epjc/s10052-016-3987-y}{{\em Eur. Phys. J.}
  {\bfseries C76} no.~4, (2016) 184},
\href{http://arxiv.org/abs/1509.08258}{{\ttfamily arXiv:1509.08258 [hep-ex]}}.

\bibitem{Aad:2012dlq}
{\bfseries ATLAS} Collaboration, G.~Aad {\em et~al.}, ``{Measurement of Upsilon
  production in 7 TeV pp collisions at ATLAS},''
  \href{http://dx.doi.org/10.1103/PhysRevD.87.052004}{{\em Phys. Rev.}
  {\bfseries D87} no.~5, (2013) 052004},
\href{http://arxiv.org/abs/1211.7255}{{\ttfamily arXiv:1211.7255 [hep-ex]}}.

\bibitem{Einhorn:1975ua}
M.~Einhorn and S.~Ellis, ``{Hadronic Production of the New Resonances: Probing
  Gluon Distributions},''
  \href{http://dx.doi.org/10.1103/PhysRevD.12.2007}{{\em Phys. Rev. D}
  {\bfseries 12} (1975) 2007}.

\bibitem{Gluck:1977zm}
M.~Gluck, J.~F. Owens, and E.~Reya, ``{Gluon Contribution to Hadronic J/$\psi$
  Production},''
\href{http://dx.doi.org/10.1103/PhysRevD.17.2324}{{\em Phys. Rev.} {\bfseries
  D17} (1978) 2324}.

\bibitem{Baranov:2016clx}
S.~Baranov and A.~Lipatov, ``{Prompt charmonia production and polarization at
  LHC in the NRQCD with $k_T$-factorization. Part III: $J/\psi$ meson},''
  \href{http://dx.doi.org/10.1103/PhysRevD.96.034019}{{\em Phys. Rev. D}
  {\bfseries 96} no.~3, (2017) 034019},
  \href{http://arxiv.org/abs/1611.10141}{{\ttfamily arXiv:1611.10141
  [hep-ph]}}.

\bibitem{Braaten:1999qk}
E.~Braaten, B.~A. Kniehl, and J.~Lee, ``{Polarization of prompt $J/\psi$ at the
  Tevatron},'' \href{http://dx.doi.org/10.1103/PhysRevD.62.094005}{{\em Phys.
  Rev.} {\bfseries D62} (2000) 094005},
\href{http://arxiv.org/abs/hep-ph/9911436}{{\ttfamily arXiv:hep-ph/9911436
  [hep-ph]}}.

\bibitem{Faccioli:2010kd}
P.~Faccioli, C.~Lourenco, J.~Seixas, and H.~K. Wohri, ``{Towards the
  experimental clarification of quarkonium polarization},''
  \href{http://dx.doi.org/10.1140/epjc/s10052-010-1420-5}{{\em Eur. Phys. J.}
  {\bfseries C69} (2010) 657--673},
\href{http://arxiv.org/abs/1006.2738}{{\ttfamily arXiv:1006.2738 [hep-ph]}}.

\bibitem{Affolder:2000nn}
{\bfseries CDF} Collaboration, T.~Affolder {\em et~al.}, ``{Measurement of
  $J/\psi$ and $\psi(2S)$ polarization in $p\bar{p}$ collisions at $\sqrt{s} =
  1.8$ TeV},'' \href{http://dx.doi.org/10.1103/PhysRevLett.85.2886}{{\em Phys.
  Rev. Lett.} {\bfseries 85} (2000) 2886--2891},
\href{http://arxiv.org/abs/hep-ex/0004027}{{\ttfamily arXiv:hep-ex/0004027
  [hep-ex]}}.

\bibitem{Abulencia:2007us}
{\bfseries CDF} Collaboration, A.~Abulencia {\em et~al.}, ``{Polarization of
  $J/\psi$ and $\psi(2S)$ Mesons Produced in $p \bar{p}$ Collisions at
  $\sqrt{s}$ = 1.96-TeV},''
  \href{http://dx.doi.org/10.1103/PhysRevLett.99.132001}{{\em Phys. Rev. Lett.}
  {\bfseries 99} (2007) 132001},
\href{http://arxiv.org/abs/0704.0638}{{\ttfamily arXiv:0704.0638 [hep-ex]}}.

\bibitem{Abelev:2011md}
{\bfseries ALICE} Collaboration, B.~Abelev {\em et~al.}, ``{$J/\psi$
  polarization in $pp$ collisions at $\sqrt{s}=7$ TeV},''
  \href{http://dx.doi.org/10.1103/PhysRevLett.108.082001}{{\em Phys. Rev.
  Lett.} {\bfseries 108} (2012) 082001},
\href{http://arxiv.org/abs/1111.1630}{{\ttfamily arXiv:1111.1630 [hep-ex]}}.

\bibitem{Acharya:2018uww}
{\bfseries ALICE} Collaboration, S.~Acharya {\em et~al.}, ``{Measurement of the
  inclusive J/ $\psi $ polarization at forward rapidity in pp collisions at
  $\mathbf {\sqrt{s} = 8}$ TeV},''
  \href{http://dx.doi.org/10.1140/epjc/s10052-018-6027-2}{{\em Eur. Phys. J.}
  {\bfseries C78} no.~7, (2018) 562},
\href{http://arxiv.org/abs/1805.04374}{{\ttfamily arXiv:1805.04374 [hep-ex]}}.

\bibitem{Aaij:2013nlm}
{\bfseries LHCb} Collaboration, R.~Aaij {\em et~al.}, ``{Measurement of
  $J/\psi$ polarization in $pp$ collisions at $\sqrt{s}=7$ TeV},''
  \href{http://dx.doi.org/10.1140/epjc/s10052-013-2631-3}{{\em Eur. Phys. J.}
  {\bfseries C73} no.~11, (2013) 2631},
\href{http://arxiv.org/abs/1307.6379}{{\ttfamily arXiv:1307.6379 [hep-ex]}}.

\bibitem{Chatrchyan:2013cla}
{\bfseries CMS} Collaboration, S.~Chatrchyan {\em et~al.}, ``{Measurement of
  the Prompt $J/\psi$ and $\psi$(2S) Polarizations in $pp$ Collisions at
  $\sqrt{s}$ = 7 TeV},''
  \href{http://dx.doi.org/10.1016/j.physletb.2013.10.055}{{\em Phys. Lett.}
  {\bfseries B727} (2013) 381--402},
\href{http://arxiv.org/abs/1307.6070}{{\ttfamily arXiv:1307.6070 [hep-ex]}}.

\bibitem{Butenschoen:2010rq}
M.~Butenschoen and B.~A. Kniehl, ``{Reconciling $J/\psi$ production at HERA,
  RHIC, Tevatron, and LHC with NRQCD factorization at next-to-leading order},''
  \href{http://dx.doi.org/10.1103/PhysRevLett.106.022003}{{\em Phys. Rev.
  Lett.} {\bfseries 106} (2011) 022003},
\href{http://arxiv.org/abs/1009.5662}{{\ttfamily arXiv:1009.5662 [hep-ph]}}.

\bibitem{Ma:2010jj}
Y.-Q. Ma, K.~Wang, and K.-T. Chao, ``{A complete NLO calculation of the
  $J/\psi$ and $\psi'$ production at hadron colliders},''
  \href{http://dx.doi.org/10.1103/PhysRevD.84.114001}{{\em Phys. Rev.}
  {\bfseries D84} (2011) 114001},
\href{http://arxiv.org/abs/1012.1030}{{\ttfamily arXiv:1012.1030 [hep-ph]}}.

\bibitem{Butenschoen:2012px}
M.~Butenschoen and B.~A. Kniehl, ``{J/$\psi$ polarization at Tevatron and LHC:
  Nonrelativistic-QCD factorization at the crossroads},''
  \href{http://dx.doi.org/10.1103/PhysRevLett.108.172002}{{\em Phys. Rev.
  Lett.} {\bfseries 108} (2012) 172002},
\href{http://arxiv.org/abs/1201.1872}{{\ttfamily arXiv:1201.1872 [hep-ph]}}.

\bibitem{Chao:2012iv}
K.-T. Chao, Y.-Q. Ma, H.-S. Shao, K.~Wang, and Y.-J. Zhang, ``{$J/\psi$
  Polarization at Hadron Colliders in Nonrelativistic QCD},''
  \href{http://dx.doi.org/10.1103/PhysRevLett.108.242004}{{\em Phys. Rev.
  Lett.} {\bfseries 108} (2012) 242004},
\href{http://arxiv.org/abs/1201.2675}{{\ttfamily arXiv:1201.2675 [hep-ph]}}.

\bibitem{Gong:2012ug}
B.~Gong, L.-P. Wan, J.-X. Wang, and H.-F. Zhang, ``{Polarization for Prompt
  J/$\psi$ and $\psi(2s)$ Production at the Tevatron and LHC},''
  \href{http://dx.doi.org/10.1103/PhysRevLett.110.042002}{{\em Phys. Rev.
  Lett.} {\bfseries 110} no.~4, (2013) 042002},
\href{http://arxiv.org/abs/1205.6682}{{\ttfamily arXiv:1205.6682 [hep-ph]}}.

\bibitem{Feng:2018ukp}
Y.~Feng, B.~Gong, C.-H. Chang, and J.-X. Wang, ``{Remaining parts of the
  long-standing $J/\psi$ polarization puzzle},''
  \href{http://dx.doi.org/10.1103/PhysRevD.99.014044}{{\em Phys. Rev.}
  {\bfseries D99} no.~1, (2019) 014044},
\href{http://arxiv.org/abs/1810.08989}{{\ttfamily arXiv:1810.08989 [hep-ph]}}.

\bibitem{Ma:2018qvc}
Y.-Q. Ma, T.~Stebel, and R.~Venugopalan, ``{$J/\psi$ polarization in the
  CGC+NRQCD approach},'' \href{http://dx.doi.org/10.1007/JHEP12(2018)057}{{\em
  JHEP} {\bfseries 12} (2018) 057},
\href{http://arxiv.org/abs/1809.03573}{{\ttfamily arXiv:1809.03573 [hep-ph]}}.

\bibitem{Chatrchyan:2012woa}
{\bfseries CMS} Collaboration, S.~Chatrchyan {\em et~al.}, ``{Measurement of
  the $\Upsilon\text{(1S)}$, $\Upsilon\text{(2S)}$ and $\Upsilon\text{(3S)}$
  polarizations in $pp$ collisions at $\sqrt{s}=7$ TeV},''
  \href{http://dx.doi.org/10.1103/PhysRevLett.110.081802}{{\em Phys. Rev.
  Lett.} {\bfseries 110} no.~8, (2013) 081802},
\href{http://arxiv.org/abs/1209.2922}{{\ttfamily arXiv:1209.2922 [hep-ex]}}.

\bibitem{Khachatryan:2016vxr}
{\bfseries CMS} Collaboration, V.~Khachatryan {\em et~al.},
  ``{$\Upsilon(\mathrm{nS})$ polarizations versus particle multiplicity in pp
  collisions at $\sqrt{s} =$ 7 TeV},''
  \href{http://dx.doi.org/10.1016/j.physletb.2016.07.065}{{\em Phys. Lett. B}
  {\bfseries 761} (2016) 31--52},
  \href{http://arxiv.org/abs/1603.02913}{{\ttfamily arXiv:1603.02913
  [hep-ex]}}.

\bibitem{Aaij:2017egv}
{\bfseries LHCb} Collaboration, R.~Aaij {\em et~al.}, ``{Measurement of the
  $\Upsilon$ polarizations in $pp$ collisions at $\sqrt{s}=7$ and 8 TeV},''
  \href{http://dx.doi.org/10.1007/JHEP12(2017)110}{{\em JHEP} {\bfseries 12}
  (2017) 110},
\href{http://arxiv.org/abs/1709.01301}{{\ttfamily arXiv:1709.01301 [hep-ex]}}.

\bibitem{Han:2014kxa}
H.~Han, Y.-Q. Ma, C.~Meng, H.-S. Shao, Y.-J. Zhang, and K.-T. Chao,
  ``{$\Upsilon(nS)$ and $\chi_b(nP)$ production at hadron colliders in
  nonrelativistic QCD},''
  \href{http://dx.doi.org/10.1103/PhysRevD.94.014028}{{\em Phys. Rev.}
  {\bfseries D94} no.~1, (2016) 014028},
\href{http://arxiv.org/abs/1410.8537}{{\ttfamily arXiv:1410.8537 [hep-ph]}}.

\bibitem{BraunMunzinger:2008tz}
P.~Braun-Munzinger and J.~Wambach, ``{The Phase Diagram of Strongly-Interacting
  Matter},'' \href{http://dx.doi.org/10.1103/RevModPhys.81.1031}{{\em Rev. Mod.
  Phys.} {\bfseries 81} (2009) 1031--1050},
\href{http://arxiv.org/abs/0801.4256}{{\ttfamily arXiv:0801.4256 [hep-ph]}}.

\bibitem{Braun-Munzinger:2015hba}
P.~Braun-Munzinger, V.~Koch, T.~Schäfer, and J.~Stachel, ``{Properties of hot
  and dense matter from relativistic heavy ion collisions},''
  \href{http://dx.doi.org/10.1016/j.physrep.2015.12.003}{{\em Phys. Rept.}
  {\bfseries 621} (2016) 76--126},
\href{http://arxiv.org/abs/1510.00442}{{\ttfamily arXiv:1510.00442 [nucl-th]}}.

\bibitem{Matsui:1986dk}
T.~Matsui and H.~Satz, ``{J/$\psi$ Suppression by Quark-Gluon Plasma
  Formation},''
\href{http://dx.doi.org/10.1016/0370-2693(86)91404-8}{{\em Phys. Lett.}
  {\bfseries B178} (1986) 416--422}.

\bibitem{Laine:2007gj}
M.~Laine, ``{A Resummed perturbative estimate for the quarkonium spectral
  function in hot QCD},''
  \href{http://dx.doi.org/10.1088/1126-6708/2007/05/028}{{\em JHEP} {\bfseries
  05} (2007) 028},
\href{http://arxiv.org/abs/0704.1720}{{\ttfamily arXiv:0704.1720 [hep-ph]}}.

\bibitem{BraunMunzinger:2000px}
P.~Braun-Munzinger and J.~Stachel, ``{(Non)thermal aspects of charmonium
  production and a new look at J/$\psi$ suppression},''
  \href{http://dx.doi.org/10.1016/S0370-2693(00)00991-6}{{\em Phys. Lett.}
  {\bfseries B490} (2000) 196--202},
\href{http://arxiv.org/abs/nucl-th/0007059}{{\ttfamily arXiv:nucl-th/0007059
  [nucl-th]}}.

\bibitem{Thews:2000rj}
R.~L. Thews, M.~Schroedter, and J.~Rafelski, ``{Enhanced $J/\psi$ production in
  deconfined quark matter},''
  \href{http://dx.doi.org/10.1103/PhysRevC.63.054905}{{\em Phys. Rev.}
  {\bfseries C63} (2001) 054905},
\href{http://arxiv.org/abs/hep-ph/0007323}{{\ttfamily arXiv:hep-ph/0007323
  [hep-ph]}}.

\bibitem{Ioffe:2003rd}
B.~L. Ioffe and D.~E. Kharzeev, ``{Quarkonium polarization in heavy ion
  collisions as a possible signature of the quark gluon plasma},''
  \href{http://dx.doi.org/10.1103/PhysRevC.68.061902}{{\em Phys. Rev.}
  {\bfseries C68} (2003) 061902},
\href{http://arxiv.org/abs/hep-ph/0306176}{{\ttfamily arXiv:hep-ph/0306176
  [hep-ph]}}.

\bibitem{Shao:2014fca}
H.-S. Shao, Y.-Q. Ma, K.~Wang, and K.-T. Chao, ``{Polarizations of $\chi_{c1}$
  and $\chi_{c2}$ in prompt production at the LHC},''
  \href{http://dx.doi.org/10.1103/PhysRevLett.112.182003}{{\em Phys. Rev.
  Lett.} {\bfseries 112} no.~18, (2014) 182003},
\href{http://arxiv.org/abs/1402.2913}{{\ttfamily arXiv:1402.2913 [hep-ph]}}.

\bibitem{Sirunyan:2019apc}
{\bfseries CMS} Collaboration, A.~M. Sirunyan {\em et~al.}, ``{Measurement of
  the $\chi_\mathrm{c1}$ and $\chi_\mathrm{c2}$ polarizations in proton-proton
  collisions at $\sqrt{s} =$ 8 TeV},''
\href{http://arxiv.org/abs/1912.07706}{{\ttfamily arXiv:1912.07706 [hep-ex]}}.

\bibitem{Aamodt:2008zz}
{\bfseries ALICE} Collaboration, K.~Aamodt {\em et~al.}, ``{The ALICE
  experiment at the CERN LHC},''
\href{http://dx.doi.org/10.1088/1748-0221/3/08/S08002}{{\em JINST} {\bfseries
  3} (2008) S08002}.

\bibitem{Abelev:2014ffa}
{\bfseries ALICE} Collaboration, B.~Abelev {\em et~al.}, ``{Performance of the
  ALICE Experiment at the CERN LHC},''
  \href{http://dx.doi.org/10.1142/S0217751X14300440}{{\em Int. J. Mod. Phys.}
  {\bfseries A29} (2014) 1430044},
\href{http://arxiv.org/abs/1402.4476}{{\ttfamily arXiv:1402.4476 [nucl-ex]}}.

\bibitem{Aamodt:2011gj}
{\bfseries ALICE} Collaboration, K.~Aamodt {\em et~al.}, ``{Rapidity and
  transverse momentum dependence of inclusive $\jpsi$ production in $\pp$
  collisions at $\sqrt{s} = 7$ TeV},''
  \href{http://dx.doi.org/10.1016/j.physletb.2011.09.054,
  10.1016/j.physletb.2012.10.060}{{\em Phys. Lett.} {\bfseries B704} (2011)
  442},
\href{http://arxiv.org/abs/1105.0380}{{\ttfamily arXiv:1105.0380 [hep-ex]}}.

\bibitem{Abbas:2013taa}
{\bfseries ALICE} Collaboration, E.~Abbas {\em et~al.}, ``{Performance of the
  ALICE VZERO system},''
  \href{http://dx.doi.org/10.1088/1748-0221/8/10/P10016}{{\em JINST} {\bfseries
  8} (2013) P10016},
\href{http://arxiv.org/abs/1306.3130}{{\ttfamily arXiv:1306.3130 [nucl-ex]}}.

\bibitem{Aamodt:2010aa}
{\bfseries ALICE} Collaboration, K.~Aamodt {\em et~al.}, ``{Alignment of the
  ALICE Inner Tracking System with cosmic-ray tracks},''
  \href{http://dx.doi.org/10.1088/1748-0221/5/03/P03003}{{\em JINST} {\bfseries
  5} (2010) P03003},
\href{http://arxiv.org/abs/1001.0502}{{\ttfamily arXiv:1001.0502
  [physics.ins-det]}}.

\bibitem{ALICE:2012aa}
{\bfseries ALICE} Collaboration, B.~Abelev {\em et~al.}, ``{Measurement of the
  Cross Section for Electromagnetic Dissociation with Neutron Emission in Pb-Pb
  Collisions at $\sqrt{s_{_{\rm NN}}}$ = 2.76 TeV},''
  \href{http://dx.doi.org/10.1103/PhysRevLett.109.252302}{{\em Phys. Rev.
  Lett.} {\bfseries 109} (2012) 252302},
\href{http://arxiv.org/abs/1203.2436}{{\ttfamily arXiv:1203.2436 [nucl-ex]}}.

\bibitem{Bossu:2012jt}
{\bfseries ALICE} Collaboration, F.~Bossu, M.~Gagliardi, and M.~Marchisone,
  ``{Performance of the RPC-based ALICE muon trigger system at the LHC},''
  \href{http://dx.doi.org/10.22323/1.159.0059,
  10.1088/1748-0221/7/12/T12002}{{\em JINST} {\bfseries 7} (2012) T12002},
  \href{http://arxiv.org/abs/1211.1948}{{\ttfamily arXiv:1211.1948
  [physics.ins-det]}}.
[PoSRPC2012,059(2012)].

\bibitem{Abelev:2013qoq}
{\bfseries ALICE} Collaboration, B.~Abelev {\em et~al.}, ``{Centrality
  determination of Pb-Pb collisions at $\sqrt{s_{_{\rm NN}}}$ = 2.76 TeV with
  ALICE},'' \href{http://dx.doi.org/10.1103/PhysRevC.88.044909}{{\em Phys.
  Rev.} {\bfseries C88} no.~4, (2013) 044909},
\href{http://arxiv.org/abs/1301.4361}{{\ttfamily arXiv:1301.4361 [nucl-ex]}}.

\bibitem{ALICE-PUBLIC-2018-011}
{\bfseries ALICE} Collaboration, ``{Centrality determination in heavy ion
  collisions},'' ALICE-PUBLIC-2018-011.
  \url{http://cds.cern.ch/record/2636623}.

\bibitem{ALICE-PUBLIC-2015-006}
{\bfseries ALICE} Collaboration, ``{Quarkonium signal extraction in ALICE},''
  ALICE-PUBLIC-2015-006. \url{https://cds.cern.ch/record/2060096}.

\bibitem{Acharya:2019iur}
{\bfseries ALICE} Collaboration, S.~Acharya {\em et~al.}, ``{Studies of
  J/$\psi$ production at forward rapidity in Pb-Pb collisions at
  $\sqrt{s_{\rm{NN}}}$ = 5.02 TeV},''
  \href{http://dx.doi.org/10.1007/JHEP02(2020)041}{{\em JHEP} {\bfseries 02}
  (2020) 041},
\href{http://arxiv.org/abs/1909.03158}{{\ttfamily arXiv:1909.03158 [nucl-ex]}}.

\bibitem{Acharya:2018kxc}
{\bfseries ALICE} Collaboration, S.~Acharya {\em et~al.}, ``{Inclusive J/$\psi$
  production at forward and backward rapidity in p-Pb collisions at
  $\sqrt{s_{\rm NN}}$ = 8.16 TeV},''
  \href{http://dx.doi.org/10.1007/JHEP07(2018)160}{{\em JHEP} {\bfseries 07}
  (2018) 160},
\href{http://arxiv.org/abs/1805.04381}{{\ttfamily arXiv:1805.04381 [nucl-ex]}}.

\bibitem{Brun:1082634}
R.~Brun, F.~Bruyant, F.~Carminati, S.~Giani, M.~Maire, A.~McPherson,
  G.~Patrick, and L.~Urban, {\em {GEANT: Detector Description and Simulation
  Tool; Oct 1994}}.
\newblock CERN Program Library. CERN, Geneva, 1993.
\newblock \url{http://cds.cern.ch/record/1082634}.
\newblock Long Writeup W5013.

\bibitem{Abelev:2014qha}
{\bfseries ALICE} Collaboration, B.~Abelev {\em et~al.}, ``{Measurement of
  quarkonium production at forward rapidity in $pp$ collisions at $\sqrt{s} =
  7$ TeV},'' \href{http://dx.doi.org/10.1140/epjc/s10052-014-2974-4}{{\em Eur.
  Phys. J.} {\bfseries C74} no.~8, (2014) 2974},
\href{http://arxiv.org/abs/1403.3648}{{\ttfamily arXiv:1403.3648 [nucl-ex]}}.

\bibitem{Acharya:2018mni}
{\bfseries ALICE} Collaboration, S.~Acharya {\em et~al.}, ``{$\Upsilon$
  suppression at forward rapidity in Pb-Pb collisions at $\sqrt{s_{\rm NN}}$ =
  5.02 TeV},'' \href{http://dx.doi.org/10.1016/j.physletb.2018.11.067}{{\em
  Phys. Lett.} {\bfseries B790} (2019) 89--101},
\href{http://arxiv.org/abs/1805.04387}{{\ttfamily arXiv:1805.04387 [nucl-ex]}}.

\end{thebibliography}\endgroup

\newpage
\appendix
\section{The ALICE Collaboration}
\label{app:collab}

\begingroup
\small
\begin{flushleft}
S.~Acharya\Irefn{org141}\And 
D.~Adamov\'{a}\Irefn{org95}\And 
A.~Adler\Irefn{org74}\And 
J.~Adolfsson\Irefn{org81}\And 
M.M.~Aggarwal\Irefn{org100}\And 
G.~Aglieri Rinella\Irefn{org34}\And 
M.~Agnello\Irefn{org30}\And 
N.~Agrawal\Irefn{org10}\textsuperscript{,}\Irefn{org54}\And 
Z.~Ahammed\Irefn{org141}\And 
S.~Ahmad\Irefn{org16}
% ...
\And
\renewcommand\labelenumi{\textsuperscript{\theenumi}~}

\section*{Affiliation notes}
\renewcommand\theenumi{\roman{enumi}}
\begin{Authlist}
\item \Adef{org*}Deceased
\item \Adef{orgI}Italian National Agency for New Technologies, Energy and Sustainable Economic Development (ENEA), Bologna, Italy
\item \Adef{orgII}Dipartimento DET del Politecnico di Torino, Turin, Italy
\item \Adef{orgIII}M.V. Lomonosov Moscow State University, D.V. Skobeltsyn Institute of Nuclear, Physics, Moscow, Russia
\item \Adef{orgIV}Department of Applied Physics, Aligarh Muslim University, Aligarh, India
\item \Adef{orgV}Institute of Theoretical Physics, University of Wroclaw, Poland
\end{Authlist}

\section*{Collaboration Institutes}
\renewcommand\theenumi{\arabic{enumi}~}
\begin{Authlist}
\item \Idef{org1}A.I. Alikhanyan National Science Laboratory (Yerevan Physics Institute) Foundation, Yerevan, Armenia
\item \Idef{org2}Bogolyubov Institute for Theoretical Physics, National Academy of Sciences of Ukraine, Kiev, Ukraine
\item \Idef{org3}Bose Institute, Department of Physics  and Centre for Astroparticle Physics and Space Science (CAPSS), Kolkata, India
\item \Idef{org4}Budker Institute for Nuclear Physics, Novosibirsk, Russia
\item \Idef{org5}California Polytechnic State University, San Luis Obispo, California, United States
\item \Idef{org6}Central China Normal University, Wuhan, China
\item \Idef{org7}Centre de Calcul de l'IN2P3, Villeurbanne, Lyon, France
\item \Idef{org8}Centro de Aplicaciones Tecnol\'{o}gicas y Desarrollo Nuclear (CEADEN), Havana, Cuba
\item \Idef{org9}Centro de Investigaci\'{o}n y de Estudios Avanzados (CINVESTAV), Mexico City and M\'{e}rida, Mexico
\item \Idef{org10}Centro Fermi - Museo Storico della Fisica e Centro Studi e Ricerche ``Enrico Fermi', Rome, Italy
\item \Idef{org11}Chicago State University, Chicago, Illinois, United States
\item \Idef{org12}China Institute of Atomic Energy, Beijing, China
\item \Idef{org13}Comenius University Bratislava, Faculty of Mathematics, Physics and Informatics, Bratislava, Slovakia
\item \Idef{org14}COMSATS University Islamabad, Islamabad, Pakistan
\item \Idef{org15}Creighton University, Omaha, Nebraska, United States
\item \Idef{org16}Department of Physics, Aligarh Muslim University, Aligarh, India
\item \Idef{org17}Department of Physics, Pusan National University, Pusan, Republic of Korea
\item \Idef{org18}Department of Physics, Sejong University, Seoul, Republic of Korea
\item \Idef{org19}Department of Physics, University of California, Berkeley, California, United States
\item \Idef{org20}Department of Physics, University of Oslo, Oslo, Norway
\item \Idef{org21}Department of Physics and Technology, University of Bergen, Bergen, Norway
\item \Idef{org22}Dipartimento di Fisica dell'Universit\`{a} 'La Sapienza' and Sezione INFN, Rome, Italy
\item \Idef{org23}Dipartimento di Fisica dell'Universit\`{a} and Sezione INFN, Cagliari, Italy
\item \Idef{org24}Dipartimento di Fisica dell'Universit\`{a} and Sezione INFN, Trieste, Italy
\item \Idef{org25}Dipartimento di Fisica dell'Universit\`{a} and Sezione INFN, Turin, Italy
\item \Idef{org26}Dipartimento di Fisica e Astronomia dell'Universit\`{a} and Sezione INFN, Bologna, Italy
\item \Idef{org27}Dipartimento di Fisica e Astronomia dell'Universit\`{a} and Sezione INFN, Catania, Italy
\item \Idef{org28}Dipartimento di Fisica e Astronomia dell'Universit\`{a} and Sezione INFN, Padova, Italy
\item \Idef{org29}Dipartimento di Fisica `E.R.~Caianiello' dell'Universit\`{a} and Gruppo Collegato INFN, Salerno, Italy
\item \Idef{org30}Dipartimento DISAT del Politecnico and Sezione INFN, Turin, Italy
\item \Idef{org31}Dipartimento di Scienze e Innovazione Tecnologica dell'Universit\`{a} del Piemonte Orientale and INFN Sezione di Torino, Alessandria, Italy
\item \Idef{org32}Dipartimento di Scienze MIFT, Universit\`{a} di Messina, Messina, Italy
\item \Idef{org33}Dipartimento Interateneo di Fisica `M.~Merlin' and Sezione INFN, Bari, Italy
\item \Idef{org34}European Organization for Nuclear Research (CERN), Geneva, Switzerland
\item \Idef{org35}Faculty of Electrical Engineering, Mechanical Engineering and Naval Architecture, University of Split, Split, Croatia
\item \Idef{org36}Faculty of Engineering and Science, Western Norway University of Applied Sciences, Bergen, Norway
\item \Idef{org37}Faculty of Nuclear Sciences and Physical Engineering, Czech Technical University in Prague, Prague, Czech Republic
\item \Idef{org38}Faculty of Science, P.J.~\v{S}af\'{a}rik University, Ko\v{s}ice, Slovakia
\item \Idef{org39}Frankfurt Institute for Advanced Studies, Johann Wolfgang Goethe-Universit\"{a}t Frankfurt, Frankfurt, Germany
\item \Idef{org40}Fudan University, Shanghai, China
\item \Idef{org41}Gangneung-Wonju National University, Gangneung, Republic of Korea
\item \Idef{org42}Gauhati University, Department of Physics, Guwahati, India
\item \Idef{org43}Helmholtz-Institut f\"{u}r Strahlen- und Kernphysik, Rheinische Friedrich-Wilhelms-Universit\"{a}t Bonn, Bonn, Germany
\item \Idef{org44}Helsinki Institute of Physics (HIP), Helsinki, Finland
\item \Idef{org45}High Energy Physics Group,  Universidad Aut\'{o}noma de Puebla, Puebla, Mexico
\item \Idef{org46}Hiroshima University, Hiroshima, Japan
\item \Idef{org47}Hochschule Worms, Zentrum  f\"{u}r Technologietransfer und Telekommunikation (ZTT), Worms, Germany
\item \Idef{org48}Horia Hulubei National Institute of Physics and Nuclear Engineering, Bucharest, Romania
\item \Idef{org49}Indian Institute of Technology Bombay (IIT), Mumbai, India
\item \Idef{org50}Indian Institute of Technology Indore, Indore, India
\item \Idef{org51}Indonesian Institute of Sciences, Jakarta, Indonesia
\item \Idef{org52}INFN, Laboratori Nazionali di Frascati, Frascati, Italy
\item \Idef{org53}INFN, Sezione di Bari, Bari, Italy
\item \Idef{org54}INFN, Sezione di Bologna, Bologna, Italy
\item \Idef{org55}INFN, Sezione di Cagliari, Cagliari, Italy
\item \Idef{org56}INFN, Sezione di Catania, Catania, Italy
\item \Idef{org57}INFN, Sezione di Padova, Padova, Italy
\item \Idef{org58}INFN, Sezione di Roma, Rome, Italy
\item \Idef{org59}INFN, Sezione di Torino, Turin, Italy
\item \Idef{org60}INFN, Sezione di Trieste, Trieste, Italy
\item \Idef{org61}Inha University, Incheon, Republic of Korea
\item \Idef{org62}Institute for Nuclear Research, Academy of Sciences, Moscow, Russia
\item \Idef{org63}Institute for Subatomic Physics, Utrecht University/Nikhef, Utrecht, Netherlands
\item \Idef{org64}Institute of Experimental Physics, Slovak Academy of Sciences, Ko\v{s}ice, Slovakia
\item \Idef{org65}Institute of Physics, Homi Bhabha National Institute, Bhubaneswar, India
\item \Idef{org66}Institute of Physics of the Czech Academy of Sciences, Prague, Czech Republic
\item \Idef{org67}Institute of Space Science (ISS), Bucharest, Romania
\item \Idef{org68}Institut f\"{u}r Kernphysik, Johann Wolfgang Goethe-Universit\"{a}t Frankfurt, Frankfurt, Germany
\item \Idef{org69}Instituto de Ciencias Nucleares, Universidad Nacional Aut\'{o}noma de M\'{e}xico, Mexico City, Mexico
\item \Idef{org70}Instituto de F\'{i}sica, Universidade Federal do Rio Grande do Sul (UFRGS), Porto Alegre, Brazil
\item \Idef{org71}Instituto de F\'{\i}sica, Universidad Nacional Aut\'{o}noma de M\'{e}xico, Mexico City, Mexico
\item \Idef{org72}iThemba LABS, National Research Foundation, Somerset West, South Africa
\item \Idef{org73}Jeonbuk National University, Jeonju, Republic of Korea
\item \Idef{org74}Johann-Wolfgang-Goethe Universit\"{a}t Frankfurt Institut f\"{u}r Informatik, Fachbereich Informatik und Mathematik, Frankfurt, Germany
\item \Idef{org75}Joint Institute for Nuclear Research (JINR), Dubna, Russia
\item \Idef{org76}Korea Institute of Science and Technology Information, Daejeon, Republic of Korea
\item \Idef{org77}KTO Karatay University, Konya, Turkey
\item \Idef{org78}Laboratoire de Physique des 2 Infinis, Ir\`{e}ne Joliot-Curie, Orsay, France
\item \Idef{org79}Laboratoire de Physique Subatomique et de Cosmologie, Universit\'{e} Grenoble-Alpes, CNRS-IN2P3, Grenoble, France
\item \Idef{org80}Lawrence Berkeley National Laboratory, Berkeley, California, United States
\item \Idef{org81}Lund University Department of Physics, Division of Particle Physics, Lund, Sweden
\item \Idef{org82}Nagasaki Institute of Applied Science, Nagasaki, Japan
\item \Idef{org83}Nara Women{'}s University (NWU), Nara, Japan
\item \Idef{org84}National and Kapodistrian University of Athens, School of Science, Department of Physics , Athens, Greece
\item \Idef{org85}National Centre for Nuclear Research, Warsaw, Poland
\item \Idef{org86}National Institute of Science Education and Research, Homi Bhabha National Institute, Jatni, India
\item \Idef{org87}National Nuclear Research Center, Baku, Azerbaijan
\item \Idef{org88}National Research Centre Kurchatov Institute, Moscow, Russia
\item \Idef{org89}Niels Bohr Institute, University of Copenhagen, Copenhagen, Denmark
\item \Idef{org90}Nikhef, National institute for subatomic physics, Amsterdam, Netherlands
\item \Idef{org91}NRC Kurchatov Institute IHEP, Protvino, Russia
\item \Idef{org92}NRC \guillemotleft Kurchatov\guillemotright~Institute - ITEP, Moscow, Russia
\item \Idef{org93}NRNU Moscow Engineering Physics Institute, Moscow, Russia
\item \Idef{org94}Nuclear Physics Group, STFC Daresbury Laboratory, Daresbury, United Kingdom
\item \Idef{org95}Nuclear Physics Institute of the Czech Academy of Sciences, \v{R}e\v{z} u Prahy, Czech Republic
\item \Idef{org96}Oak Ridge National Laboratory, Oak Ridge, Tennessee, United States
\item \Idef{org97}Ohio State University, Columbus, Ohio, United States
\item \Idef{org98}Petersburg Nuclear Physics Institute, Gatchina, Russia
\item \Idef{org99}Physics department, Faculty of science, University of Zagreb, Zagreb, Croatia
\item \Idef{org100}Physics Department, Panjab University, Chandigarh, India
\item \Idef{org101}Physics Department, University of Jammu, Jammu, India
\item \Idef{org102}Physics Department, University of Rajasthan, Jaipur, India
\item \Idef{org103}Physikalisches Institut, Eberhard-Karls-Universit\"{a}t T\"{u}bingen, T\"{u}bingen, Germany
\item \Idef{org104}Physikalisches Institut, Ruprecht-Karls-Universit\"{a}t Heidelberg, Heidelberg, Germany
\item \Idef{org105}Physik Department, Technische Universit\"{a}t M\"{u}nchen, Munich, Germany
\item \Idef{org106}Politecnico di Bari, Bari, Italy
\item \Idef{org107}Research Division and ExtreMe Matter Institute EMMI, GSI Helmholtzzentrum f\"ur Schwerionenforschung GmbH, Darmstadt, Germany
\item \Idef{org108}Rudjer Bo\v{s}kovi\'{c} Institute, Zagreb, Croatia
\item \Idef{org109}Russian Federal Nuclear Center (VNIIEF), Sarov, Russia
\item \Idef{org110}Saha Institute of Nuclear Physics, Homi Bhabha National Institute, Kolkata, India
\item \Idef{org111}School of Physics and Astronomy, University of Birmingham, Birmingham, United Kingdom
\item \Idef{org112}Secci\'{o}n F\'{\i}sica, Departamento de Ciencias, Pontificia Universidad Cat\'{o}lica del Per\'{u}, Lima, Peru
\item \Idef{org113}St. Petersburg State University, St. Petersburg, Russia
\item \Idef{org114}Stefan Meyer Institut f\"{u}r Subatomare Physik (SMI), Vienna, Austria
\item \Idef{org115}SUBATECH, IMT Atlantique, Universit\'{e} de Nantes, CNRS-IN2P3, Nantes, France
\item \Idef{org116}Suranaree University of Technology, Nakhon Ratchasima, Thailand
\item \Idef{org117}Technical University of Ko\v{s}ice, Ko\v{s}ice, Slovakia
\item \Idef{org118}The Henryk Niewodniczanski Institute of Nuclear Physics, Polish Academy of Sciences, Cracow, Poland
\item \Idef{org119}The University of Texas at Austin, Austin, Texas, United States
\item \Idef{org120}Universidad Aut\'{o}noma de Sinaloa, Culiac\'{a}n, Mexico
\item \Idef{org121}Universidade de S\~{a}o Paulo (USP), S\~{a}o Paulo, Brazil
\item \Idef{org122}Universidade Estadual de Campinas (UNICAMP), Campinas, Brazil
\item \Idef{org123}Universidade Federal do ABC, Santo Andre, Brazil
\item \Idef{org124}University of Cape Town, Cape Town, South Africa
\item \Idef{org125}University of Houston, Houston, Texas, United States
\item \Idef{org126}University of Jyv\"{a}skyl\"{a}, Jyv\"{a}skyl\"{a}, Finland
\item \Idef{org127}University of Liverpool, Liverpool, United Kingdom
\item \Idef{org128}University of Science and Technology of China, Hefei, China
\item \Idef{org129}University of South-Eastern Norway, Tonsberg, Norway
\item \Idef{org130}University of Tennessee, Knoxville, Tennessee, United States
\item \Idef{org131}University of the Witwatersrand, Johannesburg, South Africa
\item \Idef{org132}University of Tokyo, Tokyo, Japan
\item \Idef{org133}University of Tsukuba, Tsukuba, Japan
\item \Idef{org134}Universit\'{e} Clermont Auvergne, CNRS/IN2P3, LPC, Clermont-Ferrand, France
\item \Idef{org135}Universit\'{e} de Lyon, Universit\'{e} Lyon 1, CNRS/IN2P3, IPN-Lyon, Villeurbanne, Lyon, France
\item \Idef{org136}Universit\'{e} de Strasbourg, CNRS, IPHC UMR 7178, F-67000 Strasbourg, France, Strasbourg, France
\item \Idef{org137}Universit\'{e} Paris-Saclay Centre d'Etudes de Saclay (CEA), IRFU, D\'{e}partment de Physique Nucl\'{e}aire (DPhN), Saclay, France
\item \Idef{org138}Universit\`{a} degli Studi di Foggia, Foggia, Italy
\item \Idef{org139}Universit\`{a} degli Studi di Pavia, Pavia, Italy
\item \Idef{org140}Universit\`{a} di Brescia, Brescia, Italy
\item \Idef{org141}Variable Energy Cyclotron Centre, Homi Bhabha National Institute, Kolkata, India
\item \Idef{org142}Warsaw University of Technology, Warsaw, Poland
\item \Idef{org143}Wayne State University, Detroit, Michigan, United States
\item \Idef{org144}Westf\"{a}lische Wilhelms-Universit\"{a}t M\"{u}nster, Institut f\"{u}r Kernphysik, M\"{u}nster, Germany
\item \Idef{org145}Wigner Research Centre for Physics, Budapest, Hungary
\item \Idef{org146}Yale University, New Haven, Connecticut, United States
\item \Idef{org147}Yonsei University, Seoul, Republic of Korea
\end{Authlist}
\endgroup
\end{document}